\documentclass[twocolumn]{aastex62}

\shorttitle{LRG-BEASTS IV: WASP-39b}
\shortauthors{Kirk et al.}

\begin{document}

\title{LRG-BEASTS: Transmission Spectroscopy and Retrieval Analysis of the Highly-Inflated Saturn-mass Planet WASP-39b}

\correspondingauthor{James Kirk}
\email{james.kirk@cfa.harvard.edu}

\author[0000-0002-4207-6615]{James Kirk}
\affil{Center for Astrophysics $\vert$ Harvard \& Smithsonian, 60 Garden Street, Cambridge, MA 02138, USA}

\author[0000-0003-3204-8183]{Mercedes L\'{o}pez-Morales}
\affil{Center for Astrophysics $\vert$ Harvard \& Smithsonian, 60 Garden Street, Cambridge, MA 02138, USA}

\author[0000-0003-1452-2240]{Peter J. Wheatley}
\affiliation{Department of Physics, University of Warwick, Gibbet Hill Road, Coventry, CV4 7AL, UK}

\author[0000-0001-6205-6315]{Ian C. Weaver}
\affil{Center for Astrophysics $\vert$ Harvard \& Smithsonian, 60 Garden Street, Cambridge, MA 02138, USA}

\author[0000-0002-3849-8276]{Ian Skillen}
\affiliation{Isaac Newton Group of Telescopes, Apartado de Correos 321, E-38700 Santa Cruz de La Palma, Spain}

\author[0000-0002-2574-356X]{Tom Louden}
\affiliation{Department of Physics, University of Warwick, Gibbet Hill Road, Coventry, CV4 7AL, UK}

\author{James McCormac}
\affiliation{Department of Physics, University of Warwick, Gibbet Hill Road, Coventry, CV4 7AL, UK}

\author[0000-0001-9513-1449]{N\'{e}stor Espinoza}
\affiliation{Max-Planck-Institut f\"ur Astronomie, K\"onigstuhl 17, 69117 Heidelberg, Germany}



\begin{abstract}

We present a ground-based transmission spectrum and comprehensive retrieval analysis of the highly inflated Saturn-mass planet WASP-39b. We obtained low-resolution spectra ($R \approx 400$) of a transit of WASP-39b using the ACAM instrument on the 4.2\,m William Herschel Telescope as part of the LRG-BEASTS survey. Our transmission spectrum is in good agreement with previous ground- and space-based observations of WASP-39b, and covers a wavelength range of 4000--9000\,\AA. Previous analyses of this exoplanet have retrieved water abundances that span more than four orders of magnitude, which in turn lead to conclusions of a subsolar or highly supersolar atmospheric metallicity.
 In order to determine the cause of the large discrepancies in the literature regarding WASP-39b's atmospheric metallicity, we performed retrieval analyses of all literature data sets. Our retrievals, which assume equilibrium chemistry, recovered highly supersolar metallicities for all data sets. When running our retrievals on a combined spectrum, spanning 0.3--5\,$\mu$m, we recovered an atmospheric metallicity of $282^{+65}_{-58} \times$ solar. We find that stellar activity has a negligible effect on the derived abundances and instead conclude that different assumptions made during retrieval analyses lead to the reported water abundances that differ by orders of magnitude. This in turn has significant consequences for the conclusions we draw. This is the fourth planet to be observed as part of the LRG-BEASTS survey, which is demonstrating that 4\,m class telescopes can obtain low-resolution transmission spectra with precisions of around one atmospheric scale height. 

\end{abstract}



\section{Introduction} \label{sec:intro}

In the 17 years since the first detection of an exoplanet atmosphere \citep{Charbonneau2002}, the atmospheres of dozens of exoplanets have been characterized in transmission. These observations have revealed a remarkable diversity across the exoplanet population, from (relatively) small exoplanets with cloudy atmospheres (e.g.,\ \citealp{kreidberg2014_gj12}) to large exoplanets with clear atmospheres (e.g.,\ \citealp{Nikolov2018_w96}). A recent survey of 10 hot Jupiters highlighted this diversity by revealing a continuum from clear to cloudy atmospheres \citep{Sing2016}. 

Clouds and hazes are problematic for atmospheric studies as they mute, or sometimes entirely mask, absorption lines of interest (e.g., \citealp{Pont2008}; \citealp{Mallonn2016,Kirk2017,Louden2017}). In particular, they can mask features originating at altitudes lower than the aerosols themselves, such as the pressure-broadened wings of Na and K. Indeed, the pressure-broadened wings have only been observed in two exoplanets to date (WASP-39b, \citealt{Fischer2016,Sing2016}; and WASP-96b, \citealt{Nikolov2018_w96}), serving to highlight the near-ubiquitous nature of clouds and hazes.

Given the impact aerosols have on atmospheric characterization, it would be useful to know \emph{a priori} whether a planet is likely to be clear or cloudy. For this reason, there have been a number of recent studies looking for correlations between planetary parameters and the presence of clouds and hazes \citep{Heng2016,Sing2016,Stevenson2016,Crossfield2017,Fu2017}. While these studies have been conducted over relatively small sample sizes, there is growing evidence that hotter planets are more likely than cooler planets to be cloud free (e.g.,~WASP-121b, \citealp{Evans2018}; KELT-9b, \citealp{Hoeijmakers2018_kelt9,Hoeijmakers2019}). This might be due, in part, to their temperatures being too high for many species to condense. Also, at hotter temperatures, the main carbon-bearing species is CO and not CH$_4$, the photodissociation of which is responsible for soot-like hazes (e.g.,\ \citealp{Morley2015}). However, theory predicts that clouds \citep{Wakeford2015, Wakeford2017_clouds} and sulfur hazes \citep{Zahnle2009b} can exist even in the hottest exoplanets.

In addition to the effects of clouds and hazes described above, they can also lead to steeper slopes than expected from H$_2$ scattering (e.g.,\ \citealp{Pont2013}). The gradient of the slope can be used to infer information about the condensates responsible (e.g., \citealp{Wakeford2015}; \citealp{Pinhas2017}) and be used to determine the mean molecular weight of the atmosphere (e.g.,\ \citealp{Benneke2012}). The mean molecular weight obtained from the optical slope can then be used to break a degeneracy between the volume mixing ratios of gases derived from infrared data and the scale height of the atmosphere.

Optical data are also useful to provide a reference pressure, which can lead to significant improvements in the precisions of abundances derived from infrared data (e.g.,\ \citealp{Griffith2014,Wakeford2018,Pinhas2019}) and is necessary for a unique constraint on the mixing ratios of absorbers and inactive gases \citep{Benneke2012}. Indeed, in a retrieval study of 10 hot Jupiters, \cite{Pinhas2019} found that optical data were essential to obtain a reliable constraint on the abundances derived from infrared data. For the case of HD\,209458b, they found that the combination of optical data with infrared data led to a $3 \times$ better constraint on the abundance of water than infrared data alone, as the optical data were able to lift the degeneracy between reference pressure and abundance (e.g., \citealp{etangs2008_hd189}; \citealp{Griffith2014}; \citealp{Heng2017}), as the absolute number densities of species cannot be calculated without an absolute pressure scale \citep{Benneke2012}. Not only does the inclusion of optical data improve the constraint on water abundance, but it can actually lead to a different measured abundance of water (see Figure 7 of \citealp{Pinhas2019}). 

The study of clouds and hazes and the importance of optical data on precise atmospheric characterization necessitates further observations of planets at optical wavelengths. This is the principal motivation behind the Low-Resolution Ground-Based Exoplanet Atmosphere Survey using Transmission Spectroscopy (LRG-BEASTS; `large beasts') project. LRG-BEASTS is pioneering the use of 4\,m class telescopes for low-resolution transmission spectroscopy and has demonstrated that such facilities can obtain transmission spectra with precisions of 1 atmospheric scale height \citep{Kirk2017,Kirk2018,Louden2017}. Such studies are often limited by systematic and not photon noise, and so larger apertures do not necessarily provide greater precision. To date, LRG-BEASTS has revealed a Rayleigh scattering haze in the atmosphere of HAT-P-18b \citep{Kirk2017}, a gray cloud deck in the atmosphere of WASP-52b \citep{Kirk2016,Louden2017}, and a haze in the atmosphere of WASP-80b \citep{Kirk2018}. All of these planets are relatively cool ($T\mathrm{_{eq}} < 1315$\,K) and our detection of clouds and hazes is in agreement with the tentative correlation between the presence of aerosols and equilibrium temperature. 

Other recent transmission spectroscopy highlights include the first detection of He in an exoplanet atmosphere (that of the evaporating planet WASP-107b; \citealp{Spake2018}), and high-resolution observations revealing metals such as atomic Fe and Ti in the atmosphere of an ultrahot Jupiter (KELT-9b; \citealt{Hoeijmakers2018_kelt9,Hoeijmakers2019}). While transit observations give information about the limb of an exoplanet's atmosphere at pressures of millibars to microbars, observations during the secondary eclipse provide information at pressures of bars to millibars and can provide information about atmospheric species complementary to information gathered during transit. Recent highlights of secondary eclipse observations include a well-constrained metallicity measurement of the hot Jupiter WASP-18b, allowing for comparison with metallicity measurements made via transmission spectroscopy \citep{Arcangeli2018}. Finally, phase curve observations can provide information about atmospheric circulation and heat transport. Recently, HAT-P-7b was observed to show a shift in the location of its hottest point, indicating changing weather patterns within the planet's atmosphere \citep{Armstrong2016}.

\subsection{WASP-39b}

WASP-39b \citep{Faedi2011} is a highly inflated Saturn-mass planet, with a mass of 0.28\,$M_\mathrm{J}$ and a radius of 1.2\,$R_\mathrm{J}$. It orbits its G8 host with a period of 4.05 days and has an equilibrium temperature of 1116\,K \citep{Faedi2011}. This temperature puts it at the boundary at which CO takes over from CH$_4$ as the main carbon-bearing species, which occurs at $\sim$1100\,K, potentially removing CH$_4$-derived aerosols from its atmosphere \citep{Morley2015}. When considering the aerosol--temperature correlations mentioned above, it is difficult to predict \emph{a priori} whether its atmosphere is more likely to be clear or cloudy. Regardless of this, WASP-39b is an excellent target for transmission spectroscopy owing to its large atmospheric scale height (983\,km) and transit depth per scale height of $452$\,ppm. For this reason, there have been previous studies of WASP-39b's transmission spectrum. 

\emph{Hubble Space Telescope} (\emph{HST})/STIS observations of WASP-39b detected Rayleigh scattering in addition to the pressure-broadened wings of both sodium and potassium \citep{Fischer2016,Sing2016}. This transmission spectrum could be described by either a clear, H$_2$-dominated atmosphere at solar or subsolar metallicities, or a solar metallicity atmosphere with a weak haze layer \citep{Fischer2016}. This was the first time that the pressure-broadened wings of both alkali lines had been detected in an exoplanet atmosphere, as these had previously always been masked by aerosols at low pressures.

In a retrieval analysis of the \emph{HST}/STIS and \emph{Spitzer} data presented in \cite{Sing2016}, \cite{Barstow2017} found that gray atmosphere models provided the best fit to WASP-39b's transmission spectrum, although clear-atmosphere and Rayleigh scattering solutions also existed.

\cite{Nikolov2016} observed WASP-39b using Very Large Telescope (VLT)/FORS in order to provide a ground-based transmission spectrum both to complement, and compare with, the \emph{HST} results of \cite{Fischer2016} and \cite{Sing2016}. The VLT transmission spectrum was in good agreement with that of \emph{HST}, finding sodium and potassium features extending over $\sim$5 and 3 atmospheric scale heights, respectively. However, while the Na detection was statistically significant, the detection of K had a significance of only 1.2$\sigma$. The VLT study found that the atmosphere was most consistent with a cloud-free solar metallicity atmosphere with either uniform absorption from large particles or $10 \times$ Rayleigh scattering from small particles. They were, however, able to rule out strong Rayleigh scattering extending across the entire optical regime and also the presence of a cloud deck. 

Recently, \cite{Wakeford2018} presented new \emph{HST}/WFC3 measurements centered around the 1.4\,$\mu$m water feature. Their precise observations and wide wavelength coverage, when combined with the \emph{HST}/STIS and VLT observations discussed above \citep{Fischer2016,Nikolov2016,Sing2016}, allowed them to constrain the planet's atmospheric metallicity to be 151$^{+48}_{-46} \times$ solar, using water as a proxy for heavy-element abundance. When plotted against the mass-metallicity relation of the solar system, \cite{Wakeford2018} showed that this metallicity is higher than expected from this trend, although they noted that there is significant scatter expected \citep{Thorngren2016}. Its metallicity led \cite{Wakeford2018} to suggest that WASP-39b formed far out in the disk in a region rich with heavy-element planetesimals.

Furthermore, \cite{Wakeford2018} found that the inclusion of optical data, in addition to their WFC3 near-infrared data, resulted in much tighter constraints on the retrieved atmospheric abundances than when just using the WFC3 data alone. In particular, the inclusion of optical data led to a $\sim$2$\times$ improvement in the abundance constraint of water and 0.5 dex improvement in the precision of [M/H]. This was an excellent demonstration of the necessity of optical data to constrain the effects of clouds and hazes on infrared data, which will become increasingly important with the launch of \emph{James Webb Space Telescope (JWST)}.

\cite{Tsiaras2018} presented their own analysis of the same \emph{HST}/WFC3 data set used in \cite{Wakeford2018}. However, \cite{Tsiaras2018} only used data from the G141 grism, covering a wavelength range of 1.1--1.6\,$\mu$m, while \cite{Wakeford2018} additionally used the G102 grism, extending their \emph{HST}/WFC3 data down to 0.8\,$\mu$m. When using the G141 data set alone, \cite{Tsiaras2018} calculated WASP-39b's ($\log _{10} $) water abundance to be $-5.94 \pm 0.61$, which is $\approx 10^{-3} \times$ solar and differed significantly with the metallicity found by \cite{Wakeford2018}.

\cite{Pinhas2018} performed retrievals on the WFC3 data provided in \cite{Tsiaras2018} in addition to the \emph{HST}/STIS and \emph{Spitzer} data presented in \cite{Sing2016}. This was the first retrieval performed on WASP-39b data that accounted for stellar activity, and \cite{Pinhas2018} found evidence for substantial heterogeneity on the host star. This is despite WASP-39 showing weak Ca\,{\footnotesize{II}} H and K emission \citep{Faedi2011,Mancini2018}, and no photometric modulation \citep{Fischer2016}. Using this retrieval code, \cite{Pinhas2019} found WASP-39b to have an H$_2$O abundance of $0.1^{+0.42}_{-0.08} \times$ solar, significantly different from the $151^{+48}_{-46} \times$ solar metallicity atmosphere found by \cite{Wakeford2018}.

\cite{Fisher2018} also performed a retrieval analysis on WASP-39b, using only the data set of \cite{Tsiaras2018}. They found a ($\log _{10} $) water abundance of $-2.3^{+0.4}_{-0.67}$, three orders of magnitude greater than \cite{Tsiaras2018} despite an identical data set being used.

This wide range of derived metallicities and the potential impact of stellar activity necessitate further study of this planet.

The paper is organized as follows. In section \ref{sec:obs} we present our WHT/ACAM observations. In sections \ref{sec:reduction} and \ref{sec:fitting} we present the reduction and fitting of our data. The results of this data analysis are presented in section \ref{sec:results} and the combined retrieval analysis is given in section \ref{sec:retrievals}. Our discussion and conclusions are presented in sections \ref{sec:discussion} and \ref{sec:conclusions}.

\section{Observations} \label{sec:obs}

We observed a single transit of WASP-39b on the night of 2016 May 3 using the low-resolution ($R \approx 400$) spectrograph ACAM \citep{Benn2008} on the 4.2\,m William Herschel Telescope (WHT). This is the same instrument as used in our previous observations of HAT-P-18b \citep{Kirk2017}, WASP-52b \citep{Louden2017} and WASP-80b \citep{Kirk2018}. 

ACAM is well suited for transmission spectroscopy owing to its wide wavelength range ($\sim$3500--9200\,\AA), wide field of view (allowing for a greater choice of comparison stars), and wide slits (negating the potential for differential slit losses between the target and comparison). We chose to use ACAM to perform transmission spectroscopy owing to its simple optical design that uses a grism, which minimizes the potential for instrument-related systematics (such as was previously seen in VLT spectra, which was caused by its Linear Atmospheric Dispersion Corrector; \citealp{Boffin2015}).

A total of 176 biases were obtained and median-combined to form a master bias. Seventy-one spectroscopic sky flats were obtained at twilight through the same slit as used in the science frames.  These were median-combined following the removal of a master bias to create a master flat. In order to remove the sky lines from the master flat, a running median was calculated for each column of the master flat using a sliding box of 5 pixels. This box width was found to remove the sky lines while retaining the pixel-to-pixel variations. Each column of the master flat was then divided by the running median along each column to remove the sky lines.  

\begin{figure*}
\centering
\includegraphics[scale=0.7]{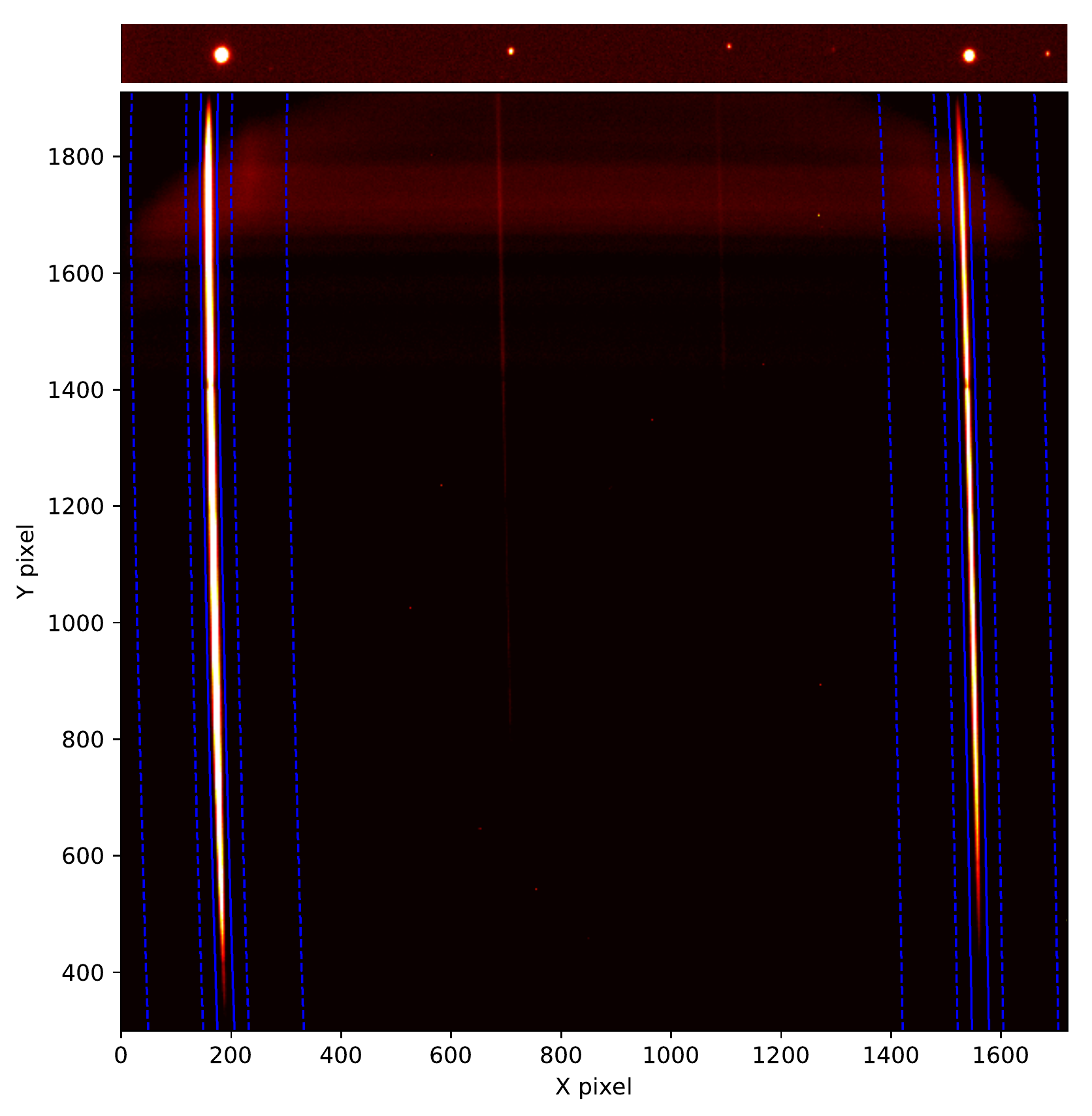}
\caption{Top panel: image taken through the 40$^{\prime\prime}$ slit. WASP-39 is the left-hand bright star, and the comparison is the right-hand bright star. This panel is cropped vertically. Bottom panel: an example science frame. WASP-39 is the left-hand bright trace, and the comparison is the right-hand bright trace. The solid blue lines show the target aperture, and the dashed lines indicate the region over which the background was calculated.}
\label{fig:extraction_frame}
\end{figure*}

\begin{figure*}
\centering
\includegraphics[scale=0.5]{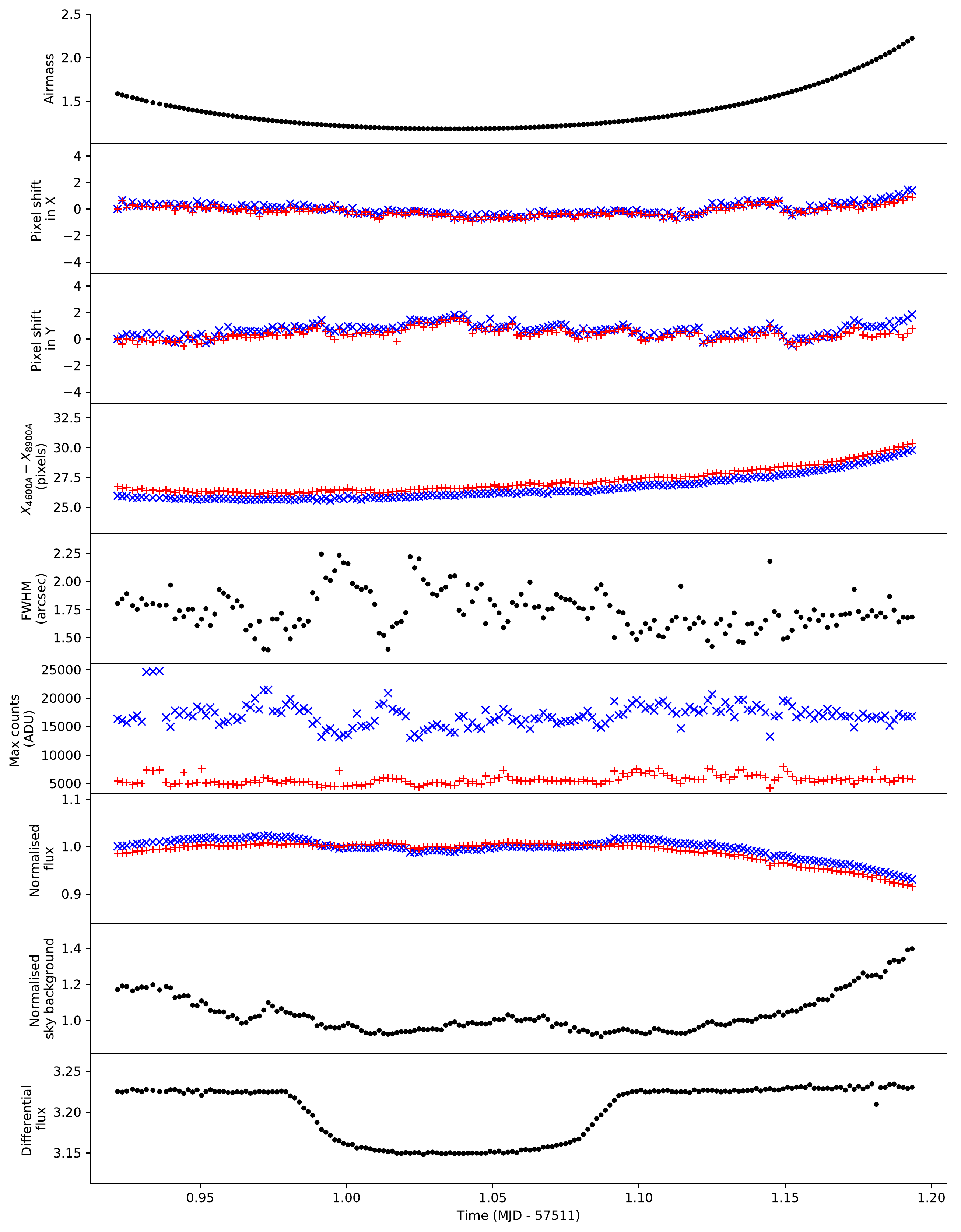}
\caption{Diagnostic plots of the night's data, all plotted with time on the $x$-axis. Top panel: variation of air mass across the observations. Second panel: the shift in the target's spectrum (blue crosses) and comparison's spectrum (red pluses) along the spatial direction. Third panel: the shift in the target's spectrum (blue crosses) and comparison's spectrum (red pluses) along the dispersion direction. Fourth panel: rotation of the target's spectrum (blue crosses) and comparison's spectrum (red pluses) as measured by the difference in the traces' locations at red and blue wavelengths. Fifth panel: the variation in FWHM across the night. Sixth panel: the maximum counts recorded in the target's spectrum (blue crosses) and comparison's spectrum (red pluses). Seventh panel: the raw white-light curves of the target (blue crosses) and comparison (red pluses). Eighth panel: the normalized sky background. Bottom panel: WASP-39's white-light curve following division by the comparison's light curve. }
\label{fig:ancillary_plots}
\end{figure*}

We obtained 179 spectra of WASP-39 and a comparison star simultaneously observed through a 40$^{\prime\prime}$ wide slit in order to remove telluric effects by performing differential photometry. The target and comparison are shown in Figure \ref{fig:extraction_frame} (top panel). Our choice of comparison star was determined by the length of the slit (7.6$^{\prime}$) and the desire to get as close a match as possible in magnitude and color to WASP-39 in order to optimize the telluric correction. The comparison star chosen was 2MASS J14292245--0321010, at a distance of 5.73$^{\prime}$ from WASP-39 with  $\Delta V = 1.1$ and $\Delta(B-V) = 0.25$ compared with WASP-39. Figure \ref{fig:ancillary_plots} shows  diagnostic plots of the night's data. 

The observations were conducted from air mass 1.58 $\rightarrow$ 1.18 
$\rightarrow$ 2.26. An exposure time of 120\,s was used, other than frames 7, 8, and 9 where a 180\,s exposure time was used. The readout time was 10\,s. We applied a small defocus and made manual guiding corrections to the telescope when the $x$ and $y$ positions of the spectral traces were observed to drift by $\gtrsim 1 $ pixel ($\approx 0.25^{\prime\prime}$; Figure \ref{fig:ancillary_plots}). The moon was not visible throughout the duration of our observations.

\section{Data reduction} \label{sec:reduction}

To reduce the data, we used our own custom-built Python scripts as introduced in \cite{Kirk2017,Kirk2018} and described in detail in \cite{Kirk2018b}. 

Following the bias and flat-field corrections using the master bias and master flat discussed in section \ref{sec:obs}, the spectral traces were extracted for the target and comparison for each of the 179 science frames. This was performed by fitting a Gaussian in the spatial direction to each trace at each row on the CCD. This gave the center of both traces as a function of $y$ (the dispersion direction). A fourth-order polynomial was then fitted to the centers calculated by the Gaussian fitting to give a smooth function of $y$. 

Following the evaluation of the target's and comparison's traces as a function of $y$, aperture photometry was then performed for each row along the dispersion direction after removal of the sky background. To calculate the sky background, a quadratic polynomial was fitted between two regions on either side of the traces and interpolated across the traces (along the spatial direction). Any stars falling within the background regions were masked from the polynomial fit as were pixels that were greater than three standard deviations from the mean (such as due to cosmic rays). Additionally, at $y$ pixels $> 1630$, ghost effects were observed in the science frames (Figure \ref{fig:extraction_frame}), which were also masked from the background aperture. 

\begin{figure}
\centering
\includegraphics[scale=0.5]{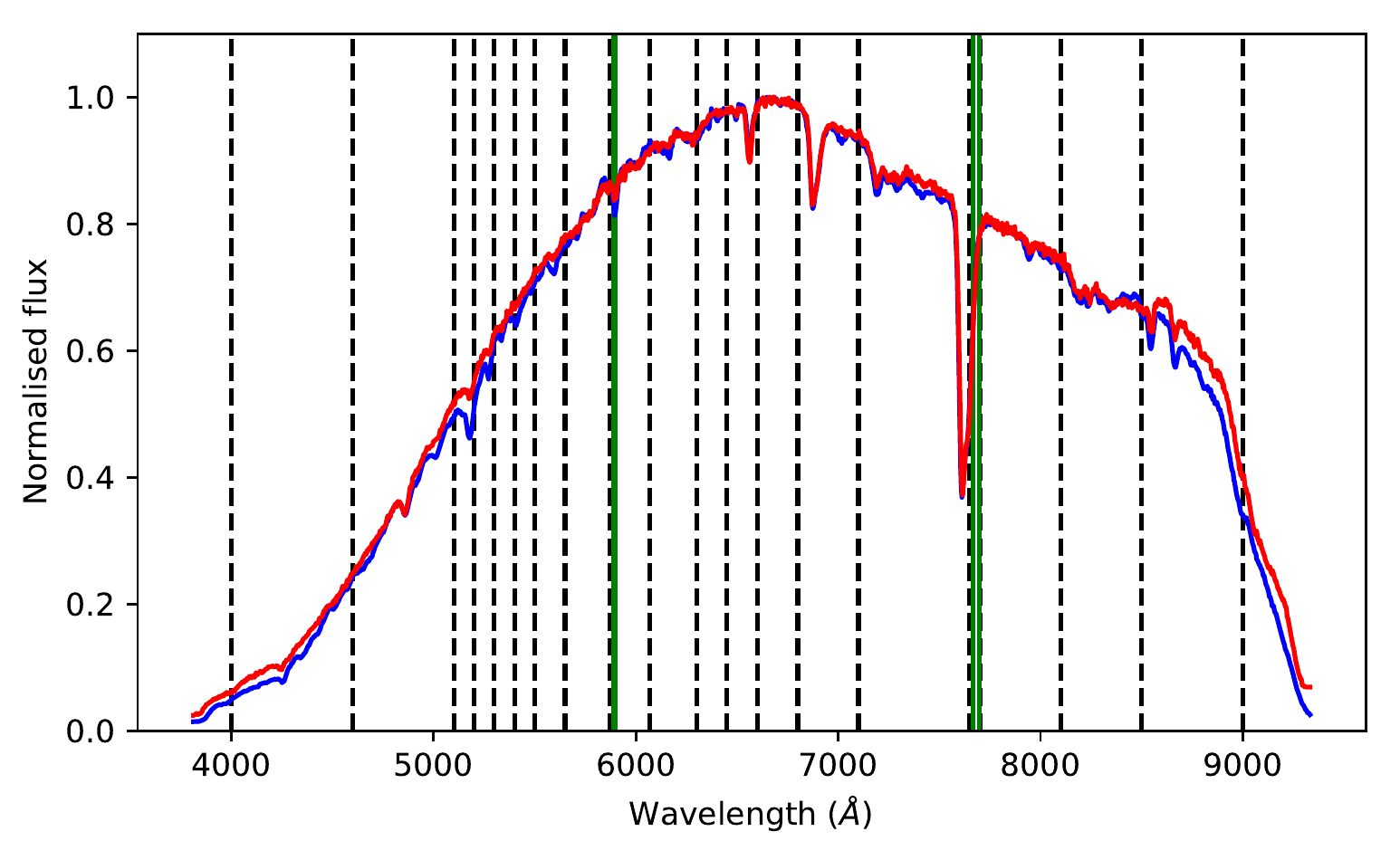}
\caption{Normalized spectra of the target (blue line) and comparison (red line). The black dashed vertical lines show the wavelength bins used to generate the spectroscopic light curves. The green vertical lines show the location of the sodium and potassium features, which are both encompassed within narrow bins (30 and 50\,\AA ~respectively). Wavelengths outside of the range 4000--9000\,\AA ~were excluded due to low signal to noise.}
\label{fig:wavelength_bins}
\end{figure}

The width of the sky background regions and apertures were experimented with in order to optimize the extraction. A 30 pixel wide aperture with two 100 pixel wide background regions on either side of each trace and offset by 26 pixels from the aperture was found to minimize the noise in the resulting white-light curve. The plate scale of ACAM is 0.253$^{\prime\prime}$\,pixel$^{-1}$. Examples of extracted spectra are shown in Figure \ref{fig:wavelength_bins}.

Following the extraction of all science spectra, cosmic rays falling within the extraction aperture were removed by linearly interpolating between the nearest neighboring unaffected pixels. The spectra of the target and comparison were then resampled onto a common wavelength grid, using a linear interpolation, so that accurate differential photometry could be performed. In order to do this, Gaussian functions were fitted to nine stellar and telluric absorption lines in both the target's and comparison's spectra and the means of these were used to describe the shifts of the absorption lines as functions of time and $y$ pixel. The wavelength solution was then calculated for the resampled spectra following the fitting of Gaussian functions to the same nine absorption lines. With the spectra resampled and a wavelength solution calculated, differential white- and spectroscopic light curves could be generated. 

Given that WASP-39b has been studied at optical wavelengths previously by both \emph{HST} \citep{Fischer2016,Sing2016} and VLT \citep{Nikolov2016}, we were guided by these binning schemes while also accounting for the differing wavelength coverage between the instruments. In total, 20 wavelength bins were used (Figure \ref{fig:wavelength_bins}) spanning a wavelength range of 4000--9000\,\AA ~with bins 3--18 corresponding exactly to the bins used by \cite{Nikolov2016} in their VLT analysis of WASP-39b. While bins 1 and 2 overlap with \cite{Nikolov2016}'s wavelength range, we opted for wider bins owing to our lower signal-to-noise ratio in the blue. Our bins 19 and 20 are redder than \cite{Nikolov2016}'s wavelength range but correspond to bins used by \cite{Fischer2016} and \cite{Sing2016}. Our white-light curve was created by integrating over the entire wavelength range.

\section{Light curve fitting} \label{sec:fitting}

With the white- and spectroscopic light curves created following the method presented in section \ref{sec:reduction}, we were then able to fit models to these to extract the parameters of interest, namely the planet-to-star radius ratio $R_\mathrm{P}/R_*$.

To fit the white-light curve, we fitted \cite{MandelAgol} analytic transit light curves with quadratic limb-darkening implemented through the \verb|Batman| Python package \citep{batman}. These were fitted simultaneously with a Gaussian process (GP) to model the systematic noise. The fitted parameters determining the transit model were the time of midtransit ($T_c$), the inclination ($i$), the ratio of semi-major axis to stellar radius ($a/R_*$), the planet-to-star radius ratio $R_\mathrm{P}/R_*$, and the linear limb-darkening coefficient ($u1$). We chose to fix the quadratic limb-darkening coefficient ($u2$) to a theoretical value as the two limb-darkening coefficients are degenerate and, for the noise levels considered in our study, there is no advantage in fitting for both coefficients \citep{Espinoza2016}. The period was held fixed to the value given in WASP-39b's discovery paper (4.055259 days; \citealt{Faedi2011}). Uniform priors were placed on all these parameters to prevent unphysical values.

To calculate the theoretical limb-darkening coefficients used in the white- and spectroscopic light curve fits we used the Limb-Darkening Toolkit (\verb|LDTk|) Python package \citep{LDTK}. This uses \textsc{phoenix} stellar atmosphere models \citep{Husser2013} with user-defined stellar parameters and uncertainties to calculate limb-darkening coefficients with uncertainties for a desired limb-darkening law. We used the stellar parameters and uncertainties listed in \cite{Faedi2011}. 

For the GP, we used the \verb|george| Python package \citep{george}. GPs are now increasingly used in the exoplanet community, from detrending \emph{K2} photometry \citep{Aigrain2016} to stellar variability analysis in radial velocity data (e.g., \citealt{Haywood2014,Rajpaul2015}) to transmission spectroscopy (e.g., \citealt{Gibson2012,Gibson2012b,Evans2015,Evans2017,Kirk2017,Kirk2018,Louden2017,Parviainen2018}). A GP is a nonparametric modeling technique that models the covariance in data. GPs are defined by kernels, which describe the  characteristics of the covariance matrix, and hyperparameters, which, in this case, define the length scale and amplitude of the covariance.

We used six squared exponential kernels taking air mass, FWHM, the mean $x$ position of the stars on the CCD, the mean $y$ positions of the stars on the CCD, the mean sky background recorded at each star's location, and time as input variables. Following the procedure of \cite{Evans2017,Evans2018}, we standardized the GP inputs by subtracting the mean and dividing by the standard deviation for each input variable. This gives each input a mean of zero and standard deviation of unity, which helps the GP determine the inputs of importance for describing the noise characteristics. The x positions and sky background were obtained as functions of the dispersion direction during the extraction process (section \ref{sec:reduction}) and therefore were binned in wavelength using the same bins as the spectra themselves. For the white-light curve, the $x$ positions and sky were integrated over the entire wavelength range. Each of these six kernels had an independent length scale ($\tau$) but shared a common amplitude ($a$). Following other studies using GPs (e.g.,\ \citealp{Evans2017,Evans2018,Gibson2017}) we fitted for the natural logarithm of these hyperparameters and fitted for the inverse length scale ($1/\tau$), each with loose uniform priors. Finally, we also included a white-noise term in the GP to account for white-noise not accounted for by the photometric error bars. This resulted in 13 fit parameters for the white light curve. 

To perform the fitting, we ran a Monte Carlo Markov chain (MCMC) using the \verb|emcee| Python package \citep{emcee}. For all our light curve fits, we began by using a running median to clip any points that deviated by $>$4$ \sigma$ from the median, which clipped at most one to two points per light curve, and then optimized the GP hyperparameters to the out-of-transit data to find the starting locations for the GP hyperparameters. The starting locations for all the transit light curve parameters were from the discovery paper \citep{Faedi2011} other than $u1$ for which we used the value calculated by \verb|LDTk|. 

For the white-light curve, we ran the MCMC for 10,000 steps with 312 walkers ($24 \times n_p$, where $n_p$ is the number of parameters) and calculated the 16th, 50th and 84th percentiles for each parameter after discarding the first 5000 steps as burn in. Following the \verb|george| documentation\footnote{https://george.readthedocs.io/en/latest/}, we then ran a second chain with the walkers initiated with a small scatter around these values for a further 10,000 steps with 432 walkers and again discarded the first 5000 steps. 

Following the fit of the white-light curve, we fitted the spectroscopic light curves but this time with $a/R_*$, $T_c$ and inclination fixed to the result from the white-light curve fit. We again fixed $u2$ to values calculated by \verb|LDTk| for our wavelength bins and fitted for $u1$ with uniform priors preventing unphysical values. This resulted in 10 fitted parameters per spectroscopic light curve. We then ran MCMCs to each light curve, following the same process as for the white-light curve but with 240 walkers as there were fewer fitted parameters.

\section{Results of WHT data analysis} \label{sec:results}

\begin{figure*}
\centering
\includegraphics[scale=1]{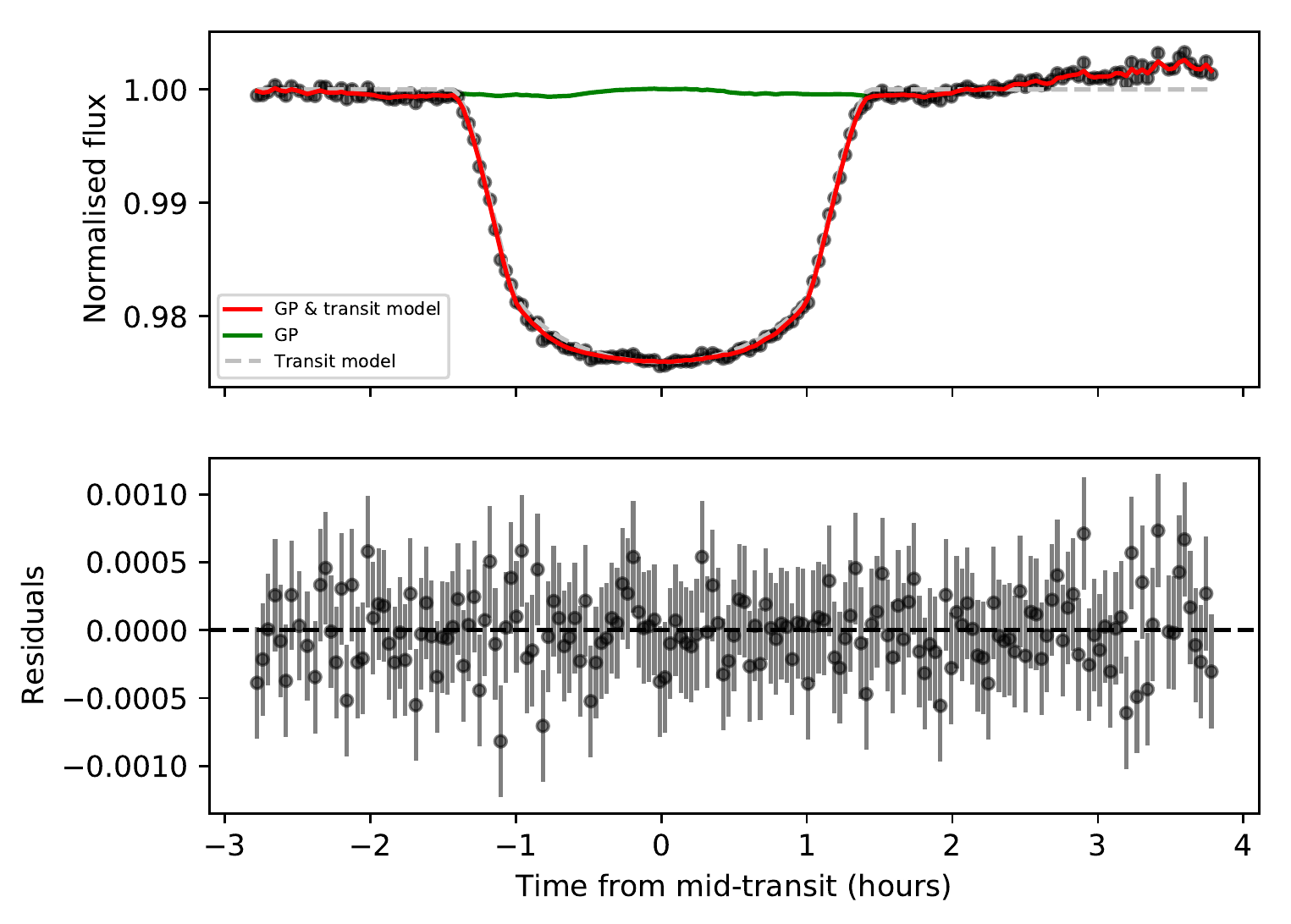}
\caption{Plot of the fit using an analytic transit light curve and Gaussian process to the white-light curve of WASP-39. Top panel: the data are shown by the black error bars with the best-fitting model shown in red. The green line and gray dashed lines show the contributions of the GP systematics model and transit light curve model respectively. Bottom panel: the residuals of the best-fitting model (red line in upper panel).}
\label{fig:white_light_fit}
\end{figure*}

The fit to the white-light curve is shown in Figure \ref{fig:white_light_fit} with the resulting parameters given in Table \ref{tab:white_light_results}, which are consistent with the literature. 

\begin{table*}
\centering
\caption{The parameters resulting from the fit to the white-light curve (Figure \ref{fig:white_light_fit}).}
\label{tab:white_light_results}
\begin{tabular}{|c|c|c|} \hline
Parameter name (units) & Symbol &  Value \\ \hline

Time of mid-transit (BJD) & $T_c$ &  $ 2457512.543848^{+0.000129}_{-0.000123} $ \\
Inclination (degrees) & $i$ & $ 87.83^{+0.2}_{-0.19} $ \\
Ratio of semi-major axis to stellar radius &  $a/R_*$ & $ 11.5^{+0.19}_{-0.21} $ \\
Ratio of planet to star radius & $R_\mathrm{P}/R_*$ & $ 0.144636^{+0.001700}_{-0.001768} $ \\
Linear limb-darkening coefficient & $u1$ & $ 0.49\pm{0.06} $ \\ 
Quadratic limb-darkening coefficient & $u2$ (fixed) & $ 0.08 $ \\ \hline

\end{tabular}

\end{table*}

\begin{figure*}
\centering
\includegraphics[scale=0.6]{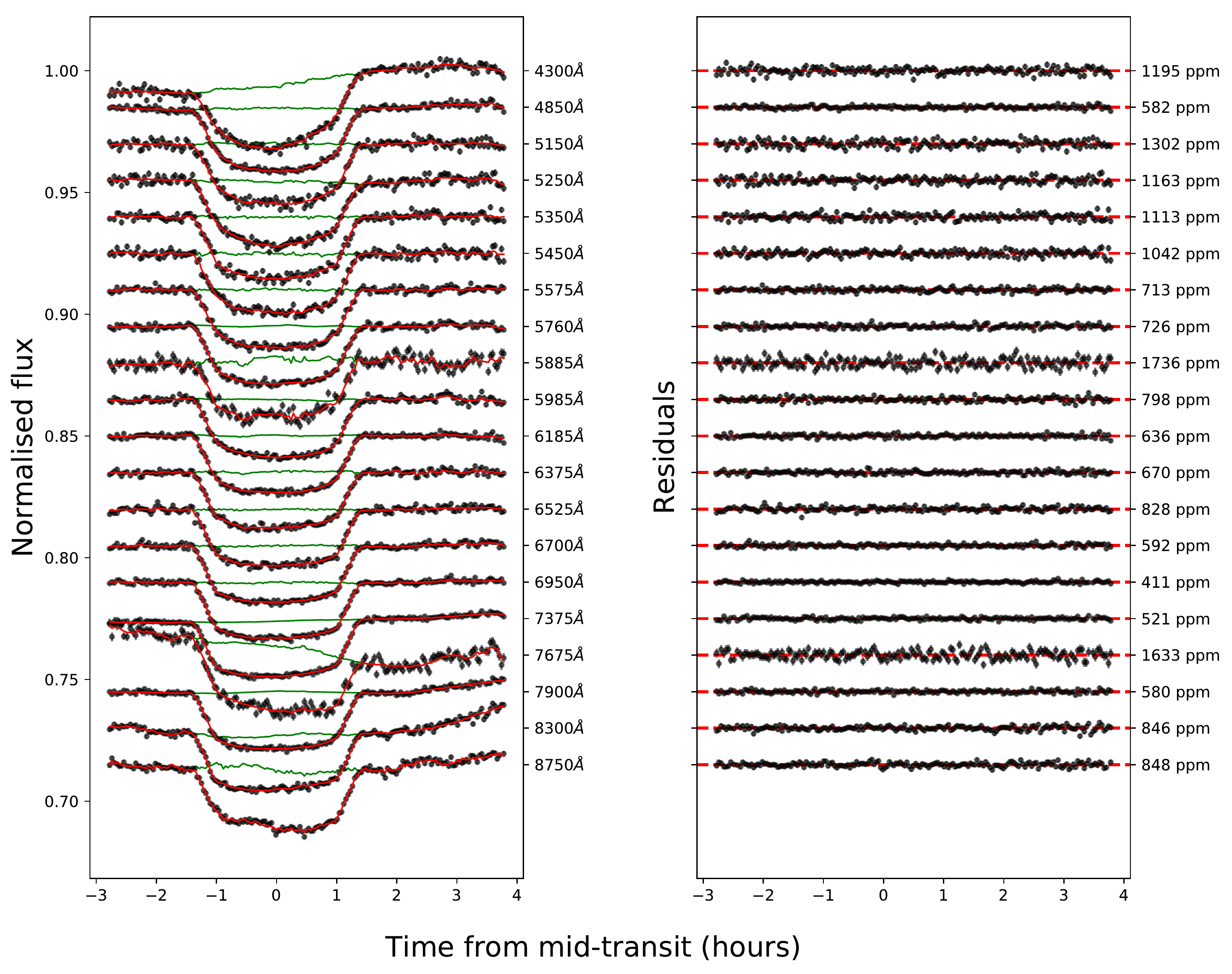}
\caption{Left-hand panel: fits to our spectroscopic light curves combining quadratically limb-darkened analytic transit light curves with a GP. The black data points show the transit light curve for each wavelength bin, with the central wavelength of each bin given on the right-hand $y$-axis, and are offset in $y$ for clarity. The green line shows the  best-fitting systematics (GP) model, and the red line shows the best-fitting systematics + analytic transit light curve model. Right-hand panel: the residuals for each fit shown in the left-hand panel, with the RMS of the residuals given on the right-hand $y$-axis. Note: the error bars are smaller than the data points.}
\label{fig:wb_fits}
\end{figure*}

The fits to the spectroscopic light curves are shown in Figure \ref{fig:wb_fits} and the resulting $R_\mathrm{P}/R_*$ values given in Table \ref{tab:wb_fit_results}.

\subsection{The transmission spectrum} \label{sec:trans_spec}

\begin{figure*}
\centering
\includegraphics[scale=0.8]{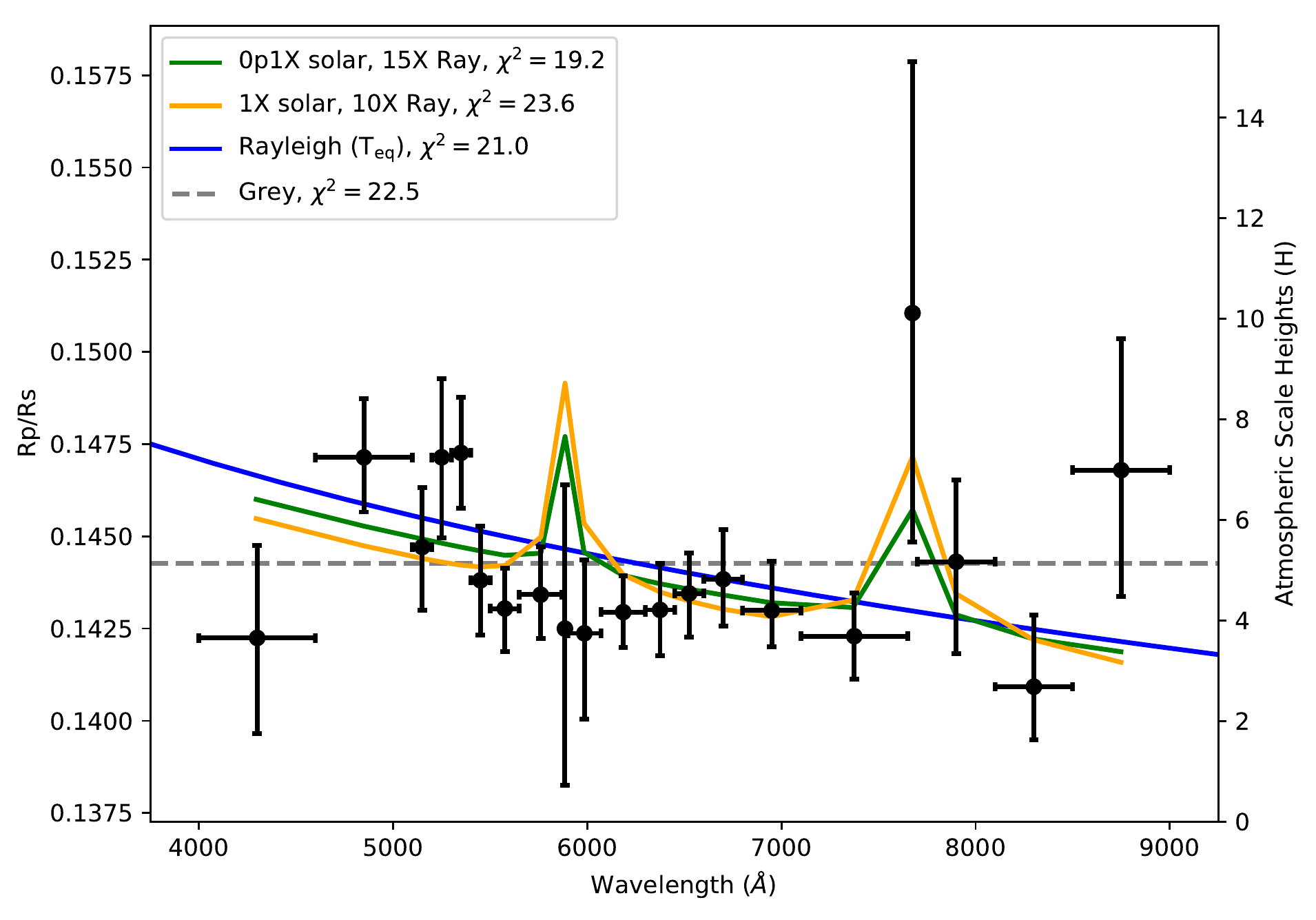}
\caption{Our WHT transmission spectrum with forward models overplotted. The green and yellow models were generated using \textsc{exo-transmit} and correspond to subsolar and solar atmospheres with Rayleigh scattering slopes enhanced by factors of 15 and 10, respectively. The blue line corresponds to a Rayleigh scattering slope at the equilibrium temperature of the planet (1116\,K; \protect\citealt{Faedi2011}). The gray dashed line shows a flat transmission spectrum as resulting from a cloud deck. Each model has 19 degrees of freedom. Note that these models have been offset in $y$ to match the mean $R_\mathrm{P}/R_*$. }
\label{fig:trans_spec}
\end{figure*}

Our transmission spectrum resulting from the fits discussed in section \ref{sec:fitting} is plotted in Figure \ref{fig:trans_spec}. This figure demonstrates our ability to achieve a precision of around one atmospheric scale height from a single transit, for wavelengths between 5500 and 7500\,\AA\ and bin widths $\geq 100$\,\AA. 

Also plotted on Figure \ref{fig:trans_spec} are two forward models generated using \textsc{exo-transmit} \citep{kempton2017}. A number of models with differing metallicities and Rayleigh scattering enhancements were generated but we only plot the best-fitting model (green line on Figure \ref{fig:trans_spec}) and the model that \cite{Nikolov2016} found best matched their data set (yellow line on Figure \ref{fig:trans_spec}). We also plot a Rayleigh scattering slope calculated using \citep{etangs2008_hd189}

\begin{equation}
\alpha H =  \frac{dR_P}{d \ln \lambda}
\end{equation} 

where $\alpha = -4$ for Rayleigh scattering and $H$ is the atmospheric scale height. Also plotted on this transmission spectrum is a flat line corresponding to a gray cloud deck.

The models shown in Figure \ref{fig:trans_spec} do not account for stellar activity, which can introduce positive and negative slopes in transmission spectra depending on whether unocculted spots or faculae are present on the stellar surface (e.g., \citealt{mccullough2014,rackham2017}). Instead, we present a retrieval analysis accounting for stellar activity in section \ref{sec:retrievals}, which shows that stellar activity plays a negligible role in the transmission spectrum of WASP-39b.

The $\chi ^2$ values for each model are shown in Figure \ref{fig:trans_spec} and indicate that the transmission spectrum favors models with scattering slopes over a flat line. However, given we do not detect significant absorption by sodium or potassium, there is little difference in the goodness of fit between models. A subsolar metallicity atmosphere with enhanced Rayleigh scattering (green line) or pure Rayleigh scattering slope (blue line) is favored, although not significantly, over the solar metallicity atmosphere with enhanced Rayleigh scattering found by \cite{Nikolov2016} to best match their VLT data. 

\begin{table*}
\centering
\caption{The tabulated transmission spectrum from the analysis of our WHT data and as plotted in Figure \protect\ref{fig:trans_spec}.}
\label{tab:wb_fit_results}
\begin{tabular}{|c|c|c|c|c|} \hline
Bin center  & Bin width & $R_\mathrm{P}/R_*$ & $u1$ & $u2$ \\
(\AA) & (\AA) & & & \\ \hline
4300   &   600  &  $ 0.14225  ^{+0.00250} _{-0.00259}  $ & $ 0.85  \pm 0.04  $ & $ -0.07  $ \\
4850   &   500  &  $ 0.14714  ^{+0.00159} _{-0.00148}  $ & $ 0.66  \pm 0.03  $ & $ 0.04  $ \\
5150   &   100  &  $ 0.14471  ^{+0.00161} _{-0.00171}  $ & $ 0.63  \pm 0.05  $ & $ 0.01  $ \\
5250   &   100  &  $ 0.14714  ^{+0.00213} _{-0.00218}  $ & $ 0.65  \pm 0.05  $ & $ 0.07  $ \\
5350   &   100  &  $ 0.14726  ^{+0.00151} _{-0.00149}  $ & $ 0.54  \pm 0.04  $ & $ 0.07  $ \\
5450   &   100  &  $ 0.14381  ^{+0.00146} _{-0.00149}  $ & $ 0.57  \pm 0.05  $ & $ 0.07  $ \\
5575   &   150  &  $ 0.14304  ^{+0.00110} _{-0.00116}  $ & $ 0.56  \pm 0.04  $ & $ 0.09  $ \\
5760   &   220  &  $ 0.14342  ^{+0.00129} _{-0.00120}  $ & $ 0.51  \pm 0.04  $ & $ 0.10  $ \\
5885   &   30  &  $ 0.14250  ^{+0.00390} _{-0.00425}  $ & $ 0.52  ^{+0.08} _{-0.09}  $ & $ 0.09  $ \\
5985   &   170  &  $ 0.14237  ^{+0.00199} _{-0.00232}  $ & $ 0.53  \pm 0.06  $ & $ 0.11  $ \\
6185   &   230  &  $ 0.14294  ^{+0.00098} _{-0.00096}  $ & $ 0.47  \pm 0.03  $ & $ 0.11  $ \\
6375   &   150  &  $ 0.14301  ^{+0.00126} _{-0.00123}  $ & $ 0.41  \pm 0.04  $ & $ 0.12  $ \\
6525   &   150  &  $ 0.14345  ^{+0.00110} _{-0.00118}  $ & $ 0.37  \pm 0.04  $ & $ 0.14  $ \\
6700   &   200  &  $ 0.14384  ^{+0.00134} _{-0.00126}  $ & $ 0.38  ^{+0.03} _{-0.04}  $ & $ 0.12  $ \\
6950   &   300  &  $ 0.14299  ^{+0.00134} _{-0.00098}  $ & $ 0.37  \pm 0.03  $ & $ 0.13  $ \\
7375   &   550  &  $ 0.14229  ^{+0.00117} _{-0.00116}  $ & $ 0.36  ^{+0.03} _{-0.02}  $ & $ 0.13  $ \\
7675   &   50  &  $ 0.15105  ^{+0.00681} _{-0.00620}  $ & $ 0.49  \pm 0.07  $ & $ 0.13  $ \\
7900   &   400  &  $ 0.14431  ^{+0.00222} _{-0.00248}  $ & $ 0.41  \pm 0.04  $ & $ 0.13  $ \\
8300   &   400  &  $ 0.14092  ^{+0.00195} _{-0.00144}  $ & $ 0.34  \pm 0.03  $ & $ 0.13  $ \\
8750   &   500  &  $ 0.14679  ^{+0.00357} _{-0.00342}  $ & $ 0.29  \pm 0.05  $ & $ 0.13  $ \\
\hline
\end{tabular}

\end{table*}

Figure \ref{fig:trans_spec} shows that the `enhancement' of the Rayleigh scattering slope serves to shift the slope up in $R_\mathrm{P}/R_*$ rather than change the gradient of the slope. This effectively decreases the amplitude of the alkali absorption lines and is why these enhanced models provide a better fit to our data in Figure \ref{fig:trans_spec}. Such an enhancement could be due to a high-altitude haze layer masking the pressure-broadened wings of the alkali wings that originate at lower altitudes. However, the amplitude of the alkali absorption lines can also be reduced by decreasing the metallicity of the atmosphere, decreasing the abundance of sodium and potassium. This introduces a degeneracy between the metallicity and the altitude of any haze layer. This is further explored in our retrieval analysis in section \ref{sec:retrievals}.

\subsection{Comparison to VLT and \emph{HST}}
\label{sec:comparison}

Because we used a binning scheme identical to that used by \cite{Nikolov2016}, we can perform a direct comparison between our WHT results and their VLT results. This binning scheme is also nearly identical to the \emph{HST}/STIS binning scheme of \cite{Fischer2016} and \cite{Sing2016}, with an offset of 10\,\AA ~between the bin edges of our bins 8, 9, 10, 17, and 18 with the corresponding \emph{HST} bins.

\begin{figure*}
\centering
\includegraphics[scale=0.7]{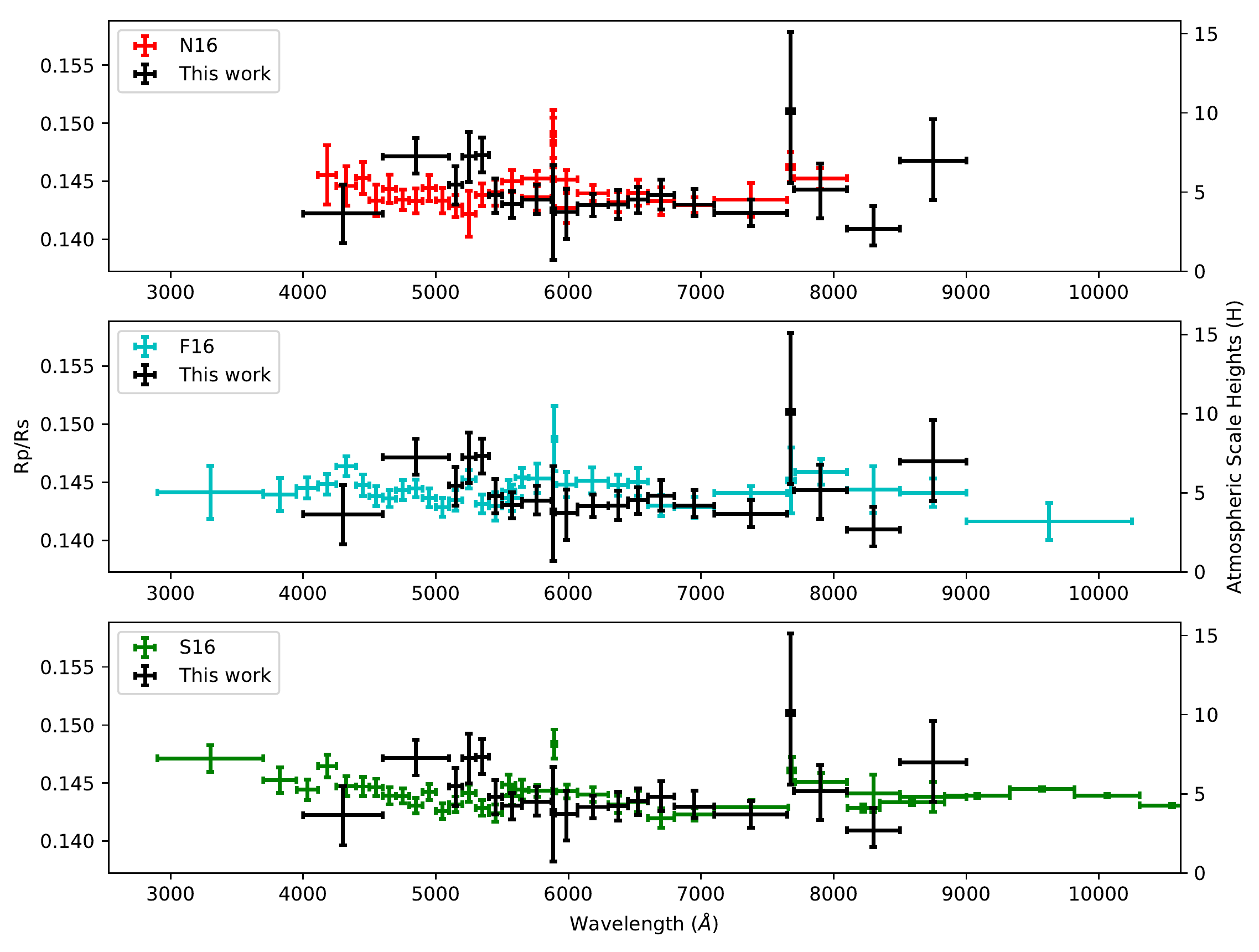}
\caption{Top panel: a comparison between the transmission spectrum resulting from our WHT/ACAM data (black error bars) and the transmission spectrum resulting from \cite{Nikolov2016}'s VLT/FORS data (red error bars). Middle panel: a comparison between the transmission spectrum resulting from our WHT/ACAM data (black error bars) and the transmission spectrum resulting from \cite{Fischer2016}'s \emph{HST}/STIS data (cyan error bars). Bottom panel: a comparison between the transmission spectrum resulting from our WHT/ACAM data (black error bars) and the transmission spectrum resulting from a combination of \cite{Sing2016}'s \emph{HST}/STIS data (green error bars) and \cite{Nikolov2016}'s VLT data as presented in \cite{Wakeford2018}. In this case, we have truncated the $x$-axis of \cite{Sing2016}'s data for clarity. Note: the VLT and \emph{HST} spectra have had small offsets applied so their means are equal to our data.}
\label{fig:comparison_plot}
\end{figure*}

Figure \ref{fig:comparison_plot} shows the comparison between our data and the studies of \cite{Nikolov2016}, \cite{Fischer2016}, and \cite{Sing2016}. However, the \cite{Sing2016} data are taken from the \cite{Wakeford2018} study and is actually is a combination of data presented in \cite{Sing2016} and \cite{Nikolov2016}. For this reason, we do not perform retrieval analysis on the \cite{Sing2016} data in section \ref{sec:retrievals} as we perform an epoch-by-epoch analysis of the stellar activity, which is lost in the combination of data.

Figure \ref{fig:comparison_plot} includes an offset applied to the VLT and \emph{HST}/STIS data sets to account for small differences between the mean $R_\mathrm{P}/R_*$ values; however, the $R_\mathrm{P}/R_*$ values are within errors between each study. This figure shows that the optical data sets are consistent with one another and demonstrates the WHT's ability to achieve precisions comparable to both the VLT and \emph{HST} over the wavelength range 5500--8500\,\AA.

\subsection{The combined transmission spectrum of WASP-39b}
\label{sec:combined_spectrum}

Here we tabulate the combined transmission spectrum of WASP-39b. We report the transmission spectrum with no offset applied to the optical data, as the mean weighted optical transmission spectrum without an offset applied is consistent with the white-light $R_\mathrm{P}/R_*$  \citep{Faedi2011}.

We chose to use \cite{Wakeford2018}'s infrared reduction owing to the overlap in reduction approaches with \cite{Fischer2016}, and the fact that it includes the \emph{HST}/WFC3 G102 (0.8 -- 1.0\,$\mu$m) grism, which the \cite{Tsiaras2018} data set does not. Our combined transmission spectrum therefore includes \emph{HST}/STIS data from \cite{Fischer2016}, VLT/FORS data from \cite{Nikolov2016}, \emph{HST}/WFC3 data from \cite{Wakeford2018}, \emph{Spitzer}/IRAC data from \cite{Sing2016}, and our WHT/ACAM optical data (Figure \ref{fig:trans_spec}). The combined transmission spectrum spans the range 0.29--5.06\,$\mu$m and is given in Table \ref{tab:combined_trans_spec}. 

\begin{table*}
\centering
\caption{The combined transmission spectrum of WASP-39b, using data from \protect\cite{Fischer2016} (F16), \protect\cite{Sing2016} (S16), \protect\cite{Nikolov2016} (N16), \protect\cite{Wakeford2018} (W18), in combination with the new optical transmission spectrum presented here (K19). The `source' column indicates which data sets contributed to the values given in the transit depth column, which are a weighted average.} \label{tab:combined_trans_spec}

\begin{tabular}{|c|c|c|c||c|c|c|c|c|c|} \hline
$\lambda$ ($\mu$m) & $\Delta \lambda$ ($\mu$m)   & $R_\mathrm{P}/R_*$ & Source & $\lambda$ ($\mu$m) & $\Delta \lambda$ ($\mu$m)   & $R_\mathrm{P}/R_*$ & Source \\ \hline
0.3355 & 0.0690 & $ 0.14429  \pm 0.00230  $ & F16 & 0.9082 & 0.0490 & $ 0.14539  \pm 0.00014  $ & W18 \\
0.3825 & 0.0250 & $ 0.14408  \pm 0.00143  $ & F16 & 0.9625 & 0.1250 & $ 0.14179  \pm 0.00161  $ & F16 \\
0.4032 & 0.0163 & $ 0.14467  \pm 0.00089  $ & F16 & 0.9572 & 0.0490 & $ 0.14598  \pm 0.00015  $ & W18 \\
0.4300 & 0.0600 & $ 0.14225  ^{+0.00250} _{-0.00259}  $ & K19 & 1.0062 & 0.0490 & $ 0.14541  \pm 0.00013  $ & W18 \\
0.4180 & 0.0140 & $ 0.14518  \pm 0.00083  $ & F16, N16 & 1.0552 & 0.0490 & $ 0.14457  \pm 0.00015  $ & W18 \\
0.4325 & 0.0150 & $ 0.14637  \pm 0.00077  $ & F16, N16 & 1.0920 & 0.0244 & $ 0.14475  \pm 0.00022  $ & W18 \\
0.4450 & 0.0100 & $ 0.14539  \pm 0.00077  $ & F16, N16 & 1.1165 & 0.0244 & $ 0.14596  \pm 0.00024  $ & W18 \\
0.4550 & 0.0100 & $ 0.14412  \pm 0.00072  $ & F16, N16 & 1.1391 & 0.0186 & $ 0.14567  \pm 0.00071  $ & W18 \\
0.4650 & 0.0100 & $ 0.14422  \pm 0.00063  $ & F16, N16 & 1.1578 & 0.0186 & $ 0.14625  \pm 0.00041  $ & W18 \\
0.4850 & 0.0500 & $ 0.14714  ^{+0.00159} _{-0.00148}  $ & K19 & 1.1765 & 0.0186 & $ 0.14611  \pm 0.00043  $ & W18 \\
0.4750 & 0.0100 & $ 0.14453  \pm 0.00062  $ & F16, N16 & 1.1951 & 0.0186 & $ 0.14542  \pm 0.00058  $ & W18 \\
0.4850 & 0.0100 & $ 0.14457  \pm 0.00062  $ & F16, N16 & 1.2138 & 0.0186 & $ 0.14500  \pm 0.00068  $ & W18 \\
0.4950 & 0.0100 & $ 0.14442  \pm 0.00065  $ & F16, N16 & 1.2325 & 0.0186 & $ 0.14536  \pm 0.00051  $ & W18 \\
0.5050 & 0.0100 & $ 0.14355  \pm 0.00065  $ & F16, N16 & 1.2512 & 0.0186 & $ 0.14576  \pm 0.00064  $ & W18 \\
0.5150 & 0.0100 & $ 0.14390  \pm 0.00061  $ & F16, N16, K19 & 1.2699 & 0.0186 & $ 0.14417  \pm 0.00040  $ & W18 \\
0.5250 & 0.0100 & $ 0.14529  \pm 0.00068  $ & F16, N16, K19 & 1.2885 & 0.0186 & $ 0.14628  \pm 0.00076  $ & W18 \\
0.5350 & 0.0100 & $ 0.14434  \pm 0.00058  $ & F16, N16, K19 & 1.3072 & 0.0186 & $ 0.14582  \pm 0.00060  $ & W18 \\
0.5450 & 0.0100 & $ 0.14416  \pm 0.00072  $ & F16, N16, K19 & 1.3259 & 0.0186 & $ 0.14663  \pm 0.00051  $ & W18 \\
0.5550 & 0.0100 & $ 0.14437  \pm 0.00098  $ & F16 & 1.3446 & 0.0186 & $ 0.14663  \pm 0.00047  $ & W18 \\
0.5575 & 0.0150 & $ 0.14467  \pm 0.00062  $ & F16, N16, K19 & 1.3633 & 0.0186 & $ 0.14687  \pm 0.00066  $ & W18 \\
0.5650 & 0.0100 & $ 0.14558  \pm 0.00083  $ & F16 & 1.3819 & 0.0186 & $ 0.14733  \pm 0.00058  $ & W18 \\
0.5765 & 0.0230 & $ 0.14561  \pm 0.00048  $ & F16, N16, K19 & 1.4006 & 0.0186 & $ 0.14749  \pm 0.00059  $ & W18 \\
0.5890 & 0.0040 & $ 0.14915  \pm 0.00125  $ & F16, N16, K19 & 1.4193 & 0.0186 & $ 0.14674  \pm 0.00050  $ & W18 \\
0.5985 & 0.0170 & $ 0.14534  \pm 0.00055  $ & F16, N16, K19 & 1.4380 & 0.0186 & $ 0.14788  \pm 0.00065  $ & W18 \\
0.6180 & 0.0240 & $ 0.14470  \pm 0.00050  $ & F16, N16, K19 & 1.4567 & 0.0186 & $ 0.14772  \pm 0.00076  $ & W18 \\
0.6375 & 0.0150 & $ 0.14438  \pm 0.00056  $ & F16, N16, K19 & 1.4753 & 0.0186 & $ 0.14803  \pm 0.00063  $ & W18 \\
0.6525 & 0.0150 & $ 0.14472  \pm 0.00067  $ & F16, N16, K19 & 1.4940 & 0.0186 & $ 0.14718  \pm 0.00068  $ & W18 \\
0.6700 & 0.0200 & $ 0.14366  \pm 0.00063  $ & F16, N16, K19 & 1.5127 & 0.0186 & $ 0.14653  \pm 0.00067  $ & W18 \\
0.6950 & 0.0300 & $ 0.14362  \pm 0.00049  $ & F16, N16, K19 & 1.5314 & 0.0186 & $ 0.14655  \pm 0.00072  $ & W18 \\
0.7375 & 0.0550 & $ 0.14401  \pm 0.00048  $ & F16, N16, K19 & 1.5501 & 0.0186 & $ 0.14656  \pm 0.00052  $ & W18 \\
0.7680 & 0.0060 & $ 0.14713  \pm 0.00116  $ & F16, N16, K19 & 1.5687 & 0.0186 & $ 0.14607  \pm 0.00068  $ & W18 \\
0.7900 & 0.0400 & $ 0.14616  \pm 0.00067  $ & F16, N16, K19 & 1.5874 & 0.0186 & $ 0.14519  \pm 0.00071  $ & W18 \\
0.8300 & 0.0400 & $ 0.14227  \pm 0.00130  $ & F16, K19 & 1.6061 & 0.0186 & $ 0.14588  \pm 0.00067  $ & W18 \\
0.8750 & 0.0500 & $ 0.14451  \pm 0.00115  $ & F16, K19 & 1.6248 & 0.0186 & $ 0.14596  \pm 0.00084  $ & W18 \\
0.8225 & 0.0244 & $ 0.14435  \pm 0.00031  $ & W18 & 1.6435 & 0.0186 & $ 0.14441  \pm 0.00061  $ & W18 \\
0.8592 & 0.0490 & $ 0.14482  \pm 0.00019  $ & W18 & 3.5600 & 0.7600 & $ 0.14438  \pm 0.00061  $ & S16 \\
& & & & 4.5000 & 1.1200 & $ 0.14659  \pm 0.00073  $ & S16 \\ \hline
\end{tabular}
\end{table*}

\section{Retrieval analysis}
\label{sec:retrievals}

In this section, we present retrieval analyses on different combinations of the data in the literature and our own data set. In particular, we wanted to test whether stellar activity could contribute to the supersolar metallicity atmosphere reported by \cite{Wakeford2018}, as \cite{Pinhas2019} report a subsolar metallicity atmosphere for WASP-39b when accounting for stellar activity in their retrievals.

To determine the effects of differing stellar activity between the epochs of \cite{Fischer2016}, \cite{Nikolov2016}, \cite{Wakeford2018}, and our observations, we ran separate retrievals on the three optical data sets (\emph{HST}, VLT, and this data), each with the \emph{HST}/WFC3 infrared data\footnote{Note: as mentioned in section \ref{sec:comparison}, we do not include the \cite{Sing2016} data set as this was taken from the \cite{Wakeford2018} paper and is actually a combination of \cite{Sing2016}'s original data and the data set of \cite{Nikolov2016}.}. We also ran retrievals on the \emph{HST}/WFC3 data presented in \cite{Tsiaras2018}, for which the authors found a subsolar water abundance. This is the same \emph{HST}/WFC3 data as used in \cite{Wakeford2018} but reduced independently.

We chose to run retrievals on the optical data and infrared data combined as the optical data are needed to constrain the effects of stellar activity while the infrared data is needed to calculate the metallicity through the abundance of water. However, we also ran retrievals on the infrared data alone to test what effect the inclusion of optical data had. 

We used the \textsc{platon} Python package \citep{Zhang2019} to perform our retrieval analysis. \textsc{platon} is a new open-source Python code, which assumes equilibrium chemistry. This algorithm has been shown by \cite{Zhang2019} to produce results consistent with another independent algorithm, \textsc{atmo} \citep{Goyal2018}, which was the algorithm used by \cite{Wakeford2018}. \textsc{platon} includes the ability to model the effects of stellar activity, parameterized by the temperature of the activity features, which can be cooler or hotter than the photosphere, and the filling factor. It then interpolates stellar atmosphere models and corrects the transit depth for the contribution of the activity region's spectrum. 

We ran retrievals on each combination of optical data (\citealp{Fischer2016,Nikolov2016} and this work) with the infrared data of \cite{Wakeford2018}. Additionally, we ran retrievals on the \cite{Wakeford2018} and \cite{Tsiaras2018} \emph{HST}/WFC3 data alone to quantify the effects of including optical data and to see what impact the differing reduction pipelines had on the retrieved parameters.

\textsc{platon} does not fit for an offset in the transit depth between data gathered at different epochs or with different instruments. We tackled this offset in two separate ways to test how sensitive our results were to the difference in the baseline transit depth between the optical and infrared data. Firstly, for retrievals run on a single optical data set plus the WFC3 data\footnote{labeled as `N16+W18', `F16+W18' and `K19+W18' in Table \ref{tab:platon_results}.}, we normalized each optical data set to the infrared data by subtracting the difference in the mean transit depth of the optical data and infrared data. This assumes that stellar activity features have a negligible effect on infrared transmission spectra, as shown by \cite{Rackham2019} for FGK dwarfs, but spots can be seen in the infrared for very active stars (e.g.,~WASP-52b; \citealt{Bruno2018}). For the combined transmission spectrum, we took the weighted mean of the three optical data sets with no offset in the transit depth applied to the optical data as the weighted mean was consistent with the white-light transit depth \citep{Faedi2011}.

Given that \textsc{platon} does not support wavelengths shorter than 3000\,\AA, we had to amend the bluest bin of \cite{Fischer2016} from 2900--3700\,\AA ~to 3010--3700\,\AA. The retrieved parameters were the planet's radius at a pressure of 1 bar ($R_\mathrm{P}$), equilibrium temperature ($T_\mathrm{eq}$), the logarithm of the atmospheric metallicity relative to the solar value ($\log Z$), and the C/O ratio. \textsc{platon} includes the treatment of clouds and hazes via the logarithm of the cloud-top pressure ($\log P_\mathrm{cloud}$), the logarithm of a factor that multiplies the scattering slope, raising the slope up and down in transit depth ($\log s$), and the gradient of the scattering slope ($\alpha$). These were all fitted parameters in our retrievals using \textsc{platon}. We also ran retrievals with each of these parameters plus the effects of stellar activity, parameterized through the active region's temperature ($T_\mathrm{active}$) and covering fraction ($f_\mathrm{active}$). 

We placed wide, uniform priors on all parameters retrieved by \textsc{platon} (Table \ref{tab:retrieval_priors}). We used the nested sampling algorithm to explore the parameter space with 1000 live points, as recommended by the documentation \citep{Zhang2019}. Depending on the data set used, the nested sampling algorithm took between $\sim$15,000 and $\sim$35,000 likelihood evaluations before convergence.

\begin{table*}
\centering
\caption{The prior ranges and parameters used for the retrievals using \textsc{platon}. Some of the parameter ranges were defined using values from the discovery paper (F11; \protect\citealp{Faedi2011}). Note that not all retrievals were run using all parameters (see Table \protect\ref{tab:platon_results} for which parameters were included for which data set).}
\label{tab:retrieval_priors}
\begin{tabular}{lcc}  \hline 
Parameter & Units & Prior \\ \hline 

Planet radius at 1 bar ($R_\mathrm{P}$) & $R_\mathrm{J}$ & Uniform ($0.9 \times R_\mathrm{P}$(F11), $1.1 \times R_\mathrm{P}$(F11))  \\
Equilibrium temperature ($T_\mathrm{eq}$) & K & Uniform ($0.5 \times T_\mathrm{eq}$(F11), $1.5 \times T_\mathrm{eq}$(F11)) \\
Metallicity ($\log Z/Z_{\odot}$) & - & Uniform (-1, 3) \\ 
C/O ratio & - & Uniform (0.05, 2.0) \\
Cloud-top pressure ($\log P_\mathrm{cloud}$) & Pa & Uniform (-0.99, 5)  \\
Scattering factor ($\log s$) & - & Uniform (-10, 10)  \\
Scattering gradient ($\alpha$) & - & Uniform (-4, 10)  \\
Unocculted spot/facula temperature ($T_\mathrm{active}$) & K & Uniform ($T_\mathrm{eff}$(F11) - 2500, $T_\mathrm{eff}$(F11) + 2500)  \\
Unocculted spot/facula covering fraction ($f_\mathrm{active}$) & - & Uniform (0, 0.5)  \\ \hline

\end{tabular}

\end{table*}

\begin{figure*}
\centering
\includegraphics[scale=0.6]{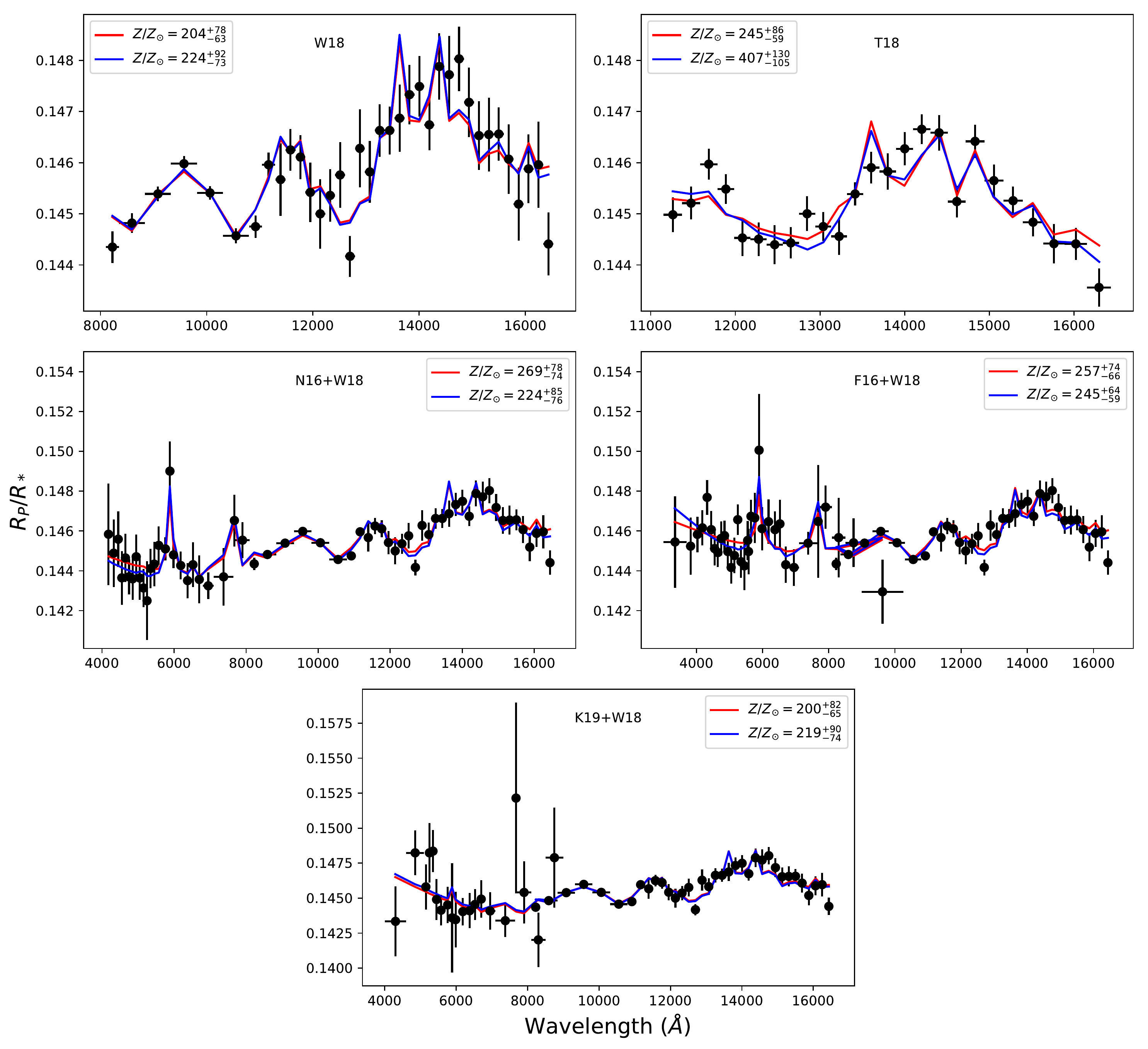}
\caption{Results from the running of \textsc{platon} retrievals. Each panel is labeled according to which combination of data sets is used. In each panel, the black points show the data, the solid red line shows the retrieved model without accounting for stellar activity, and the blue line shows the retrieval accounting for stellar activity. The legends give the atmospheric metallicity relative to solar for all models.}
\label{fig:platon_results}
\end{figure*}

\begin{figure*}
\centering
\includegraphics[scale=0.6]{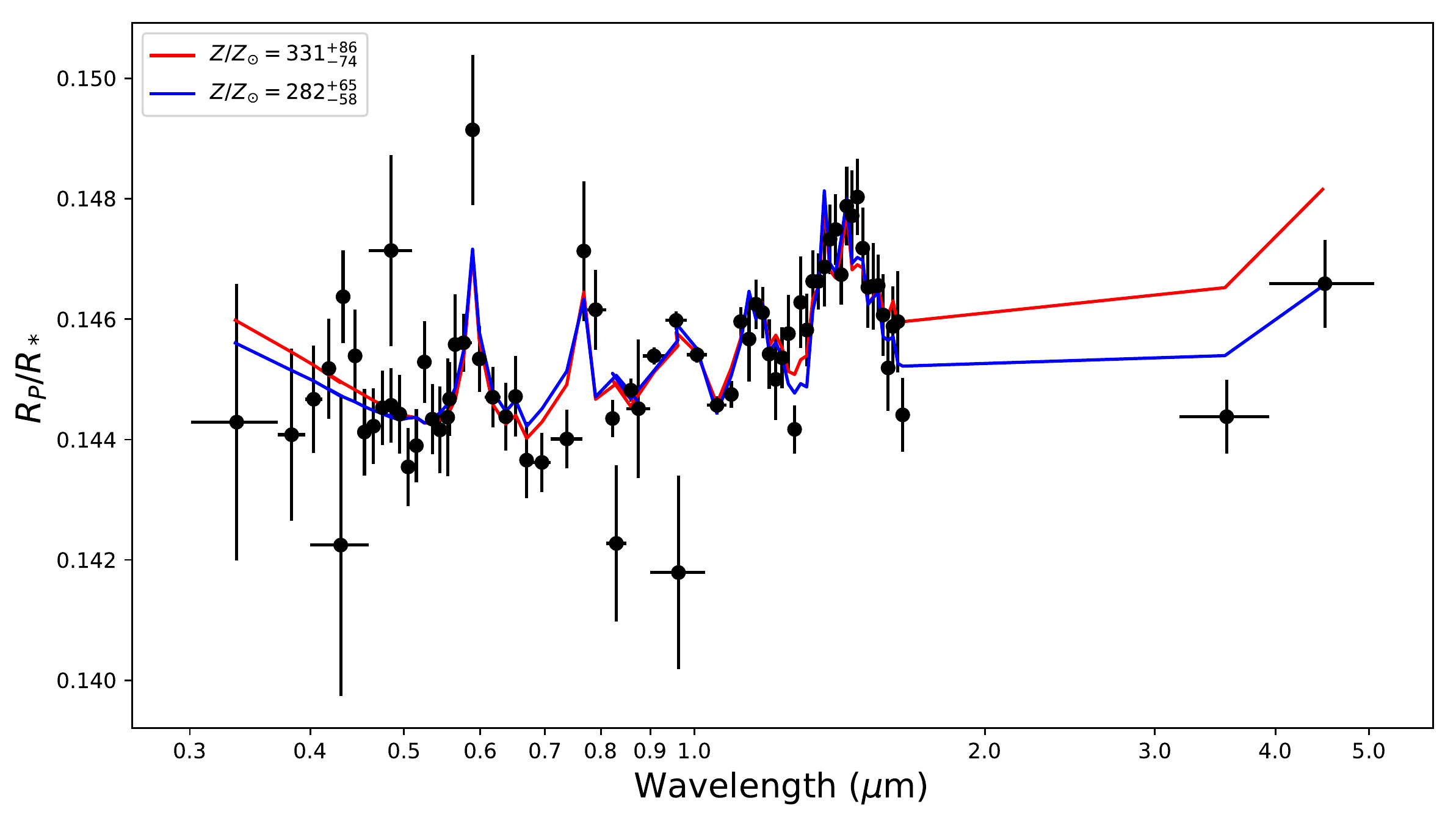}
\caption{The \textsc{platon} retrievals run on the combined transmission spectrum, incorporating data from five different studies (Table \ref{tab:combined_trans_spec}). The red line shows the retrieved model without fitting for stellar activity, and the blue line shows the retrieved model accounting for stellar activity. The legend shows the metallicity, relative to solar, retrieved by each model.}
\label{fig:platon_results_combined}
\end{figure*}

The resulting values for each retrieval using \textsc{platon} are shown in Table \ref{tab:platon_results} with retrieved model atmospheres shown in Figures \ref{fig:platon_results} and \ref{fig:platon_results_combined}. The key findings of these retrievals are that \textsc{platon} finds a highly supersolar metallicity atmosphere regardless of which data set is used, and stellar activity has a negligible effect on both the C/O ratio and the metallicity, with the exception being the retrieval performed on the \cite{Tsiaras2018} data set alone. In this case, when including stellar activity, the metallicity increases from $245^{+86}_{-59} \times$ solar to $407^{+130}_{-105} \times$ solar, although with large uncertainties. 

The fits to the combined transmission spectrum (Table \ref{tab:platon_results} and Figure \ref{fig:platon_results_combined}) retrieve a metallicity of $282^{+65}_{-58} \times$ and $331^{+86}_{-74} \times$ solar depending on whether or not stellar activity is taken into account. The nested sampling algorithm calculates the Bayesian evidences (which we call $\ln \chi$ so as not to be confused with the metallicity, $\log Z$), which marginally favor the retrieval with stellar activity with $\Delta \ln \chi = 3$.

\cite{Wakeford2018} also retrieved a supersolar metallicity atmosphere, of $151^{+48}_{-46} \times $ solar (1.2$\sigma$ lower than the value we find), while \cite{Tsiaras2018} found a subsolar water abundance. However, our retrievals on the \cite{Tsiaras2018} data alone recover a highly supersolar metallicity atmosphere, which differs significantly from the finding of these authors. Our retrieval using \textsc{platon} is also in disagreement with the results of \cite{Pinhas2019} who find a subsolar metallicity atmosphere for WASP-39b when running a retrieval on the \cite{Sing2016} and \cite{Tsiaras2018} data. These discrepancies are discussed further in section \ref{sec:metallicity}.

Furthermore, when considering Figure \ref{fig:platon_results}, it is clear that the study of \cite{Tsiaras2018} found a significantly offset and lower amplitude water feature than the study of \cite{Wakeford2018}. The reason for this difference is unclear but could be related to a different treatment of systematic noise and different system parameters between the two studies. However, despite the differing amplitudes, our retrievals on the \cite{Tsiaras2018} data using \textsc{platon} \citep{Zhang2019} also find supersolar metallicities. We considered whether our supersolar metallicities were dependent on the degeneracy between the volume mixing ratio of water, which sets the atmospheric metallicity, and both the reference pressure and radius (e.g., \citealp{etangs2008_hd189,Griffith2014,Heng2017,Pinhas2019}).

To test the effects of the cloud-top pressure and reference radius on the derived metallicities using \textsc{platon}, we also ran a retrieval on the \cite{Tsiaras2018} data alone with these parameters fixed to the values reported in that study. We find that in this case that fixing the cloud-top pressure and reference radius does not change the conclusion of a highly supersolar metallicity atmosphere (Table \ref{tab:platon_results}). We discuss this further in section \ref{sec:discrepancies}.

While the C/O ratio varies depending on which data set is used (Table \ref{tab:platon_results}), the retrieved parameters are broadly consistent with the solar C/O ratio of 0.54, when using optical and infrared data and are subsolar when using the infrared data alone. 

The lack of impact caused by stellar activity is perhaps not unexpected as both the C/O and metallicity are driven by the infrared data, which is not as affected by stellar activity, for which we only have one epoch of data. This can be seen by the consistency between the $\log Z$ values retrieved from the infrared data alone and infrared + optical data. This result implies that stellar activity is not the reason behind the supersolar metallicity atmosphere we derive for WASP-39b. This is also in agreement with the star being inactive as determined through a lack of photometric variation (\citealp{Faedi2011}; \citealp{Fischer2016}) and weak Ca\,{\footnotesize{II}} H and K emission ($\log R'_{\mathrm{HK}} = -4.97 \pm 0.06$; \citealp{Mancini2018}).

\begin{deluxetable*}{l|ccccccccc}
\tabletypesize{\scriptsize}
\tablecaption{PLATON retrieval results for the parameters as defined in the text. Retrievals were run on different combinations of \protect\cite{Wakeford2018} (W18), \protect\cite{Tsiaras2018} (T18), \protect\cite{Fischer2016} (F16), \protect\cite{Nikolov2016} (N16) and this work (K19). Note: \textsc{platon} fits for the cloud top pressure in units of Pa, but we have converted these to units of bar here. \label{tab:platon_results}}
\tablewidth{0pt}
\tablehead{
\colhead{Data set} & \colhead{$R_\mathrm{P}$ ($R_\mathrm{J}$)} & \colhead{$T_\mathrm{eq}$ (K)} & \colhead{$\log Z$} & \colhead{C/O} & \colhead{$\log P_\mathrm{cloud}$ (bar)} & \colhead{$\log s$} & \colhead{$\alpha$} & \colhead{$T_\mathrm{active}$ (K)} & \colhead{$f_\mathrm{active}$}}
\startdata
W18 w/o activity & $ 1.23 ^{+0.00} _{-0.00} $ & $ 880 ^{+99} _{-66} $ & $ 2.31 ^{+0.14} _{-0.16} $ & $ 0.32 ^{+0.20} _{-0.18} $ & $ -1.07 ^{+0.71} _{-0.86} $ & $ 0.47 ^{+0.34} _{-0.32} $ & - & - & - \\
W18 w/ activity & $ 1.23 ^{+0.01} _{-0.02} $ & $ 970 ^{+229} _{-127} $ & $ 2.35 ^{+0.15} _{-0.17} $ & $ 0.38 ^{+0.19} _{-0.22} $ & $ -1.23 ^{+0.78} _{-0.86} $ &  $ 0.49 ^{+0.34} _{-0.33} $ & - & $ 4665 ^{+758} _{-1427} $ & $ 0.04 ^{+0.07} _{-0.03} $ \\ \hline
T18 w/o activity & $ 1.21 ^{+0.01} _{-0.01} $ & $ 1180 ^{+175} _{-148} $ & $ 2.39 ^{+0.13} _{-0.12} $ & $ 0.21 ^{+0.19} _{-0.11} $ & $ -2.28 ^{+1.44} _{-1.21} $ & $ 4.02 ^{+0.45} _{-0.91} $ & - & - & - \\
T18 w/ activity & $ 1.17 ^{+0.03} _{-0.02} $ & $ 1246 ^{+221} _{-187} $ & $ 2.61 ^{+0.12} _{-0.13} $ & $ 0.26 ^{+0.22} _{-0.15} $ & $ -3.06 ^{+1.81} _{-0.75} $ & $ -2.27 ^{+4.88} _{-5.00} $ & - & $ 3478 ^{+1127} _{-370} $ & $ 0.13 ^{+0.04} _{-0.05} $ \\ \hline
T18 w/o activity & 1.24 (fixed) &  $ 893 ^{+115} _{-96} $ & $ 2.72 ^{+0.11} _{-0.12} $ & $ 0.11 ^{+0.11} _{-0.04} $ & -0.14 (fixed) & $ -3.44 ^{+4.56} _{-4.44} $ & - & - & -  \\
T18 w/ activity & 1.24 (fixed) & $ 871 ^{+113} _{-86} $ & $ 2.74 ^{+0.11} _{-0.10} $ &  $ 0.12 ^{+0.12} _{-0.05} $ & -0.14 (fixed) & $ -3.43 ^{+4.29} _{-4.32} $ & - & $ 5325 ^{+87} _{-420} $ & $ 0.06 ^{+0.20} _{-0.05} $ \\ \hline
N16+W18 w/o activity & $ 1.23 ^{+0.00} _{-0.01} $ & $ 1036 ^{+77} _{-71} $ & $ 2.43 ^{+0.11} _{-0.14} $ & $ 0.46 ^{+0.17} _{-0.22} $ & $ -1.85 ^{+0.65} _{-0.56} $ & $ -3.65 ^{+3.32} _{-4.35} $ & $ 2.87 ^{+4.66} _{-4.52} $ & - & -  \\
N16+W18 w/activity & $ 1.23 ^{+0.01} _{-0.01} $ & $ 1011 ^{+93} _{-57} $ & $ 2.35 ^{+0.14} _{-0.18} $ & $ 0.52 ^{+0.15} _{-0.28} $ & $ -1.38 ^{+0.51} _{-0.70} $ & $ -4.10 ^{+3.26} _{-3.93} $  & $ 3.20 ^{+4.41} _{-4.69} $ & $ 4422 ^{+902} _{-1177} $ & $ 0.03 ^{+0.08} _{-0.02} $ \\ \hline
F16+W18 w/o activity & $ 1.23 ^{+0.01} _{-0.01} $ & $ 1076 ^{+99} _{-111} $ & $ 2.41 ^{+0.11} _{-0.13} $ & $ 0.46 ^{+0.19} _{-0.27} $ & $ -1.12 ^{+0.75} _{-0.81} $ & $ 1.66 ^{+0.54} _{-0.79} $ & $ 4.12 ^{+1.92} _{-1.18} $ & - & - \\ 
F16+W18 w/activity & $ 1.21 ^{+0.01} _{-0.02} $ & $ 1212 ^{+122} _{-109} $ & $ 2.39 ^{+0.10} _{-0.12} $ & $ 0.55 ^{+0.11} _{-0.20} $ & $ -1.91 ^{+1.19} _{-0.65} $ & $ -0.62 ^{+1.64} _{-5.51} $ & $ 4.52 ^{+3.45} _{-4.20} $ & $ 3696 ^{+1317} _{-508} $ & $ 0.06 ^{+0.03} _{-0.02} $ \\ \hline
K19+W18 w/o activity & $ 1.23 ^{+0.00} _{-0.00} $ & $ 897 ^{+110} _{-71} $ & $ 2.30 ^{+0.15} _{-0.17} $ & $ 0.34 ^{+0.19} _{-0.18} $ & $ -1.31 ^{+0.86} _{-0.89} $ & $ 0.28 ^{+0.87} _{-1.20} $ & $ 7.07 ^{+2.08} _{-3.72} $ & - & - \\
K19+W18 w/ activity & $ 1.23 ^{+0.01} _{-0.02} $ & $ 975 ^{+219} _{-114} $ & $ 2.34 ^{+0.15} _{-0.18} $ & $ 0.35 ^{+0.20} _{-0.19} $ & $ -1.73 ^{+1.09} _{-0.86} $ & $ -0.12 ^{+1.16} _{-5.36} $ & $ 6.01 ^{+2.84} _{-5.09} $ & $ 4944 ^{+444} _{-1552} $ & $ 0.04 ^{+0.09} _{-0.03} $ \\ \hline
\textbf{Combined w/o activity} & $ 1.23 ^{+0.00} _{-0.01} $ & $ 1110 ^{+62} _{-61} $ & $ 2.52 ^{+0.10} _{-0.11} $ & $ 0.43 ^{+0.15} _{-0.21} $ & $ -2.00 ^{+1.31} _{-0.71} $ & $ -0.53 ^{+1.66} _{-5.62} $ & $ 4.51 ^{+3.55} _{-3.80} $ & - & - \\
\textbf{Combined w/ activity} & $ 1.20 ^{+0.01} _{-0.01} $ & $ 1133 ^{+108} _{-75} $ & $ 2.45 ^{+0.09} _{-0.10} $ & $ 0.25 ^{+0.19} _{-0.14} $ & $ -2.17 ^{+0.94} _{-0.42} $ & $ -2.51 ^{+2.55} _{-4.99} $ & $ 4.07 ^{+3.89} _{-4.96} $ & $ 3257 ^{+709} _{-267} $ & $ 0.05 ^{+0.02} _{-0.01} $ \\
\enddata
\end{deluxetable*}

\section{Discussion}
\label{sec:discussion}

\subsection{On the discrepancies in the literature}
\label{sec:discrepancies}

As noted throughout this work, there have been significantly different water abundances, and hence metallicities, reported for WASP-39b in the literature. The reported water abundances are given in Table \ref{tab:lit_water} and are taken from \cite{Wakeford2018}, \cite{Tsiaras2018}, \cite{Fisher2018}, and \cite{Pinhas2019}\footnote{We note that while \cite{Barstow2017} also ran retrievals on WASP-39b data, they did not have access to \emph{HST}/WFC3 observations so were unable to place tight constraints on the water abundance.}. We also include the metallicities that these water abundances correspond to, assuming a scaled solar composition atmosphere and a solar water abundance of $3.6 \times 10^{-4}$, as used by \cite{Wakeford2018}.

Table \ref{tab:lit_water} highlights how the retrieved water abundances differ by over four orders of magnitude in the literature. This includes orders-of-magnitude differences between retrievals run on identical data sets (\citealp{Fisher2018,Tsiaras2018} and Table \ref{tab:platon_results})\footnote{We also note that \cite{Fisher2018} recovered a water abundance for WASP-76b that differed at the order-of-magnitude level to \cite{Tsiaras2018} despite an identical data set being used.}. In order for us to make conclusions about the planet's atmospheric metallicity, we must first consider the causes for the large differences in the literature.

\begin{table*}
\centering
\caption{The reported logarithmic water abundances ($\log X_{\mathrm{H_2O}}$) as reported in the literature and how these compare to the solar value, along with the data sets used in the retrievals (using the same paper references as in Table \ref{tab:combined_trans_spec}).}
\label{tab:lit_water}
\begin{tabular}{l|c|c|c}
Study & Data sets used & $\log X_\mathrm{H_2O}$ & $Z/Z_{\odot}$ \\ \hline
\cite{Wakeford2018}, equilibrium chemistry  & S16, N16, W18* & - & $ 151^{+48}_{-46} $ \\
\cite{Wakeford2018}, disequilibrium chemistry & S16, N16, W18* & $-1.37^{+0.05}_{-0.13}$ & $ 117^{+14}_{-30} $ \\
\cite{Tsiaras2018} & T18** & $-5.94 \pm 0.61$ & $ 0.003^{+0.010}_{-0.002} $  \\ 
\cite{Fisher2018} & T18** & $-2.3^{+0.40}_{-0.67}$ & $ 14^{+21}_{-11} $ \\
\cite{Pinhas2019} & S16, T18** & $-4.07^{+0.72}_{-0.78}$ & $ 0.24^{+1.00}_{-0.20} $ \\
\textbf{This work} & N16, F16, S16, W18*, K19 & - & $282^{+65}_{-58} $  \\
\hline
\multicolumn{3}{l}{*using \emph{HST}/WFC3 grisms G102 and G141.} \\
\multicolumn{3}{l}{**using \emph{HST}/WFC3 grism G141 only.} \\
\end{tabular}
\end{table*}

In the retrievals used in section \ref{sec:retrievals}, we used isothermal temperature--pressure profiles. \textsc{platon} \citep{Zhang2019} has the ability to use non-isothermal temperature--pressure profiles both for forward modeling of transmission spectra and retrievals of eclipse spectra but not yet for retrievals of transmission spectra. In a retrieval analysis of simulated \emph{JWST} data, \cite{Rocchetto2016} showed that assuming an isothermal temperature--pressure profile can lead to retrieved abundances that are typically an order of magnitude too large and with underestimated uncertainties. \cite{Rocchetto2016} found that a parameterized temperature--pressure profile, however, recovered abundances that were generally within 1$\sigma$ of the true values. Conversely, \cite{Heng2017} demonstrated that isothermal temperature--pressure profiles were able to produce transmission spectra that matched numerical calculations with non-isothermal profiles, when applied to \emph{HST}/WFC3 data.

Concerning the temperature--pressure profiles used in the literature, both our retrievals and those of \cite{Tsiaras2018} assumed isothermal temperature--pressure profiles. \cite{Wakeford2018} fitted for the temperature--pressure profile in their retrieval and found that parameterized profiles were unnecessary to fit their data, instead finding the retrieved temperature--pressure profile to be isothermal. \cite{Pinhas2019} fitted for the temperature--pressure profile using the parametric profile of \cite{Madhusudhan2009} and found a non-isothermal temperature--pressure profile and a subsolar metallicity atmosphere. Finally, \cite{Fisher2018} found no strong evidence for a non-isothermal temperature--pressure profile and recovered a supersolar water abundance.

Given the lack of correlation between the treatment of the temperature--pressure profile and the water abundances reported in the literature, it is not obvious how much of an effect this has. Instead, the discrepancies could be related to a degeneracy between the reference pressure, reference radius and the abundance of water (e.g., \citealp{etangs2008_hd189,Griffith2014,Heng2017,Pinhas2019}), although \cite{Welbanks2019} find this degeneracy to have little effect on abundance estimates.

A direct comparison between the reference radii and pressures reported in the literature is more difficult. This is because these values are typically defined and treated differently for each study. However, to test the effect of this degeneracy, we also ran a retrieval on the \cite{Tsiaras2018} data alone with the reference radius and cloud-top pressure fixed to the values reported in that study. However, despite doing this, we still recover a highly supersolar metallicity (Table \ref{tab:platon_results}). Instead, we find that the equilibrium temperature recovered from this fit is lower than when fitting for the radius and pressure. The reason for this change in temperature is perhaps related to our assumption of an isothermal temperature--pressure profile.

It is also possible that the differing results in the literature result from whether equilibrium chemistry was assumed during the retrieval analysis. The retrievals of \cite{Tsiaras2018}, \cite{Fisher2018} and \cite{Pinhas2019} do not assume equilibrium chemistry and instead they retrieve for the water abundance directly. The water abundances retrieved by these studies vary from highly subsolar to marginally supersolar (Table \ref{tab:lit_water}). The retrievals presented here, however, assume equilibrium chemistry and recover highly supersolar metallicities for all literature data sets (Tables \ref{tab:platon_results} and \ref{tab:lit_water}). \cite{Wakeford2018} present the results of retrievals assuming equilibrium and free (disequilibrium) chemistry. These both result in highly supersolar metallicities of $151^{+48}_{-46} \times$ and $117^{+14}_{-30} \times$ solar (Table \ref{tab:lit_water}), respectively. This suggests that the assumption of equilibrium vs.\ disequilibrium chemistry is not the driving factor behind the discrepancies in the literature. However, we intend to perform the same analysis presented here but without the assumption of equilibrium chemistry in a future work.

Despite the above discussion that the assumption of equilibrium chemistry is unlikely to be the cause of the discrepancies, we should also consider the priors placed on the water volume mixing ratios used in the literature. \cite{Tsiaras2018}, \cite{Fisher2018}, and \cite{Pinhas2019} all used log-uniform priors, bounded by $10^{-8}$ -- $10^{-1}$, $10^{-13}$ -- 1, and $10^{-12}$ -- $10^{-2}$ respectively. \cite{Wakeford2018} do not state the prior ranges used, but the water volume mixing ratio they derive using their free chemistry model ($-1.37^{+0.05}_{-0.13}$) is excluded by the bounds of \cite{Pinhas2019}. The upper bounds are motivated by the fact that water should be a trace species in a gas-giant atmosphere (particularly one as low density as WASP-39b). However, the location of this bound should be explored when considering a planet with a large-amplitude water feature.

The above discussion highlights the impact that different assumptions and fitting methods have on the abundances of retrieved species, which are heavily model dependent.

\subsection{The atmospheric metallicity of WASP-39b}
\label{sec:metallicity}

In this section, we discuss the implications of the supersolar metallicity derived by \textsc{platon}. We are encouraged to pursue this interpretation as our analysis includes more data sets than any previous study of WASP-39b.

As detailed in section \ref{sec:retrievals}, we retrieve a supersolar metallicity atmosphere for all data sets, including the \cite{Tsiaras2018} data set. Our retrieval on the combined transmission spectrum found a $282^{+65}_{-58} \times$ solar metallicity atmosphere when taking into account stellar activity, as was marginally preferred.
 
However, the optical transmission spectra, when taken alone and when compared with forward models, are best represented by solar or even subsolar metallicities (section \ref{sec:trans_spec} and \citealp{Fischer2016,Nikolov2016}). The metallicities in the optical data are determined through the amplitude of the alkali absorption features, which are degenerate with the altitude of clouds and hazes (as discussed in section \ref{sec:trans_spec}). It is therefore possible that the optical data underestimate the actual metallicity of the atmosphere, which is better constrained by the amplitude of the water feature in the infrared. In addition, the condensation of alkali chlorides (which occurs at $\sim$800\,K, \citealp{Burrows1999}) could remove sodium and potassium from the upper, cooler atmosphere. This condensation could result in the muted alkali absorption features we see without invoking a low-metallicity atmosphere (Figure \ref{fig:trans_spec}).

Given WASP-39b's very supersolar metallicity, it sits above the trend observed for the solar system giant planets \citep{Wakeford2018}. This begs the question of how WASP-39b would accumulate such a supersolar metallicity atmosphere. In our retrievals using \textsc{platon} (section \ref{sec:retrievals} and Table \ref{tab:platon_results}), we also recovered the C/O ratios, which were broadly consistent with a solar C/O ratio. Given WASP-39 is a G star with solar metallicity \citep{Faedi2011}, we will assume that this means that WASP-39b's C/O ratio is stellar. The combination of a stellar C/O ratio and super-stellar metallicity suggests that a large amount of icy material polluted the atmosphere postformation \citep{Oberg2011}. Indeed, disk formation followed by migration can lead to $10 \times$ solar metallicities \citep{Madhusudhan2014b}, but this is still an order of magnitude less than we derive (Table \ref{tab:platon_results}).

As an additional challenge to such a metal-rich atmosphere, \cite{Thorngren2019} calculated the maximum atmospheric metallicity of WASP-39b to be $54.5 \times $ solar, significantly less than what we find and \cite{Wakeford2018} find. \cite{Thorngren2019} calculated the bulk metallicity of the planet's interior from its mass and radius combined with an evolutionary model which describes the planet's radius as a function of time \citep{Thorngren2016}. \cite{Thorngren2019} used the bulk metallicity of the interior as an upper limit to the atmospheric metallicity, as the atmosphere cannot be more metal rich than the interior postformation due to Rayleigh-Taylor instabilities and convection.

However, \cite{Thorngren2019} note that the maximum metallicity they calculate could be underestimated if the planet's interior is hotter than expected, if the planet is tidally heated, or is much younger than the models, although they predict these effects to be minimal for hot Jupiters. This makes the supersolar metallicity atmosphere of WASP-39b hard to explain, unless there is ongoing ablation of metal-rich material in its atmosphere. In any case, the atmosphere must be well mixed in order to be so rich in metals, suggesting a minimal core or composition gradient \citep{Thorngren2019}.

\section{Conclusions}
\label{sec:conclusions}

We present a new optical transmission spectrum of the Saturn-mass planet WASP-39b. Our optical transmission spectrum was acquired through the LRG-BEASTS transmission spectroscopy survey and is in good agreement with previous VLT and \emph{HST}/STIS spectra. This study further demonstrates the WHT's ability to achieve transmission spectra with precisions comparable to both \emph{HST} and VLT over a wavelength range 5500--8500\,\AA.

We performed a retrieval analysis using different combinations of all published transmission spectra of this planet. Our retrievals, which assume equilibrium chemistry, recover a highly supersolar metallicity atmosphere. This supports \cite{Wakeford2018}'s finding of a supersolar metallicity atmosphere but is at odds with the studies of \cite{Tsiaras2018}, \cite{Fisher2018} and \cite{Pinhas2019}. If the atmospheric metallicity of WASP-39b is as supersolar as our retrievals suggest, this poses a challenge to our understanding of the formation of this planet as its atmospheric metallicity is significantly enhanced relative to its interior.

We show that the orders-of-magnitude discrepancies regarding the atmospheric metallicity of WASP-39b in the literature are not due to stellar activity, as we demonstrate this to have a minimal effect. Instead, they are more likely due to differing retrieval approaches, in particular the treatment of the reference pressure and planetary radius. 

This work highlights how differing assumptions during retrieval analyses can lead to water abundances that differ by orders of magnitude, in this case over four. This is a concern as WASP-39b is one of the very best targets for transmission spectroscopy. We must explore and understand such biases before we can claim to accurately retrieve the abundances of smaller planets with \emph{JWST} data.

\section{Software and third party data repository citations} \label{sec:cite}

%

%
%

\acknowledgments

The authors thank the anonymous referee for insightful comments which improved the manuscript. The authors also wish to thank David Sing for sharing the transmission spectrum of WASP-39b, in addition to useful discussions regarding the fitting process. The authors also wish to thank Tom Evans, Rapha\"elle Haywood, and Laura Kreidberg for helpful and insightful discussions regarding Gaussian Processes and retrievals. 

%

\vspace{5mm}
\facilities{WHT (ACAM)}


\software{Astropy \citep{astropy}, Batman \citep{batman}, Emcee \citep{emcee}, George \citep{george}, LDTk \citep{LDTK}, Matplotlib \citep{matplotlib}, Numpy \citep{numpy}, Scipy \citep{scipy}}

\bibliography{combined_bib}

%
%
%



\end{document}